\documentclass[aps,pra,a4paper,preprint,unpublished]{quantumarticle}
\pdfoutput=1

\usepackage{hyperref}

\usepackage{tikz}
\usepackage{lipsum}
\usepackage{graphicx,color}
\usepackage{amsmath}
\usepackage{bm}
\usepackage{amssymb}
\usepackage{epstopdf}
\usepackage{hyperref}
\usepackage[all]{hypcap}
\usepackage{mathtools}
\usepackage{dsfont}
\usepackage{physics}
\usepackage{enumitem}
\usepackage{lipsum}
\usepackage{float}
\usepackage{tabularx}
\usepackage{listings}
\usepackage{xcolor}
\usepackage{easyReview}
\definecolor{codegreen}{rgb}{0,0.6,0}
\definecolor{codegray}{rgb}{0.5,0.5,0.5}
\definecolor{codepurple}{rgb}{0.58,0,0.82}
\definecolor{backcolour}{rgb}{0.95,0.95,0.92}

\lstdefinestyle{mystyle}{
	backgroundcolor=\color{backcolour},   
	commentstyle=\color{codegreen},
	keywordstyle=\color{magenta},
	numberstyle=\tiny\color{codegray},
	stringstyle=\color{codepurple},
	basicstyle=\ttfamily\footnotesize,
	breakatwhitespace=false,         
	breaklines=true,                 
	captionpos=b,                    
	keepspaces=true,                 
	numbers=left,                    
	numbersep=5pt,                  
	showspaces=false,                
	showstringspaces=false,
	showtabs=false,                  
	tabsize=2
}
\begin{document}
\title{Wigner State and Process Tomography on Near-Term Quantum Devices}
\author{Amit Devra}
\email{amit.devra@tum.de}
\affiliation{\footnotesize Technische Universität München, Department Chemie, Lichtenbergstrasse 4, 85747 Garching, Germany}
\affiliation{\footnotesize Munich Center for Quantum Science and Technology (MCQST), 80799 München, Germany}
\author{Niklas J. Glaser}
\affiliation{\footnotesize Munich Center for Quantum Science and Technology (MCQST), 80799 München, Germany}
\affiliation{\footnotesize Technische Universität München, Department of Physics, 85748 Garching, Germany}
\affiliation{\footnotesize Walther-Meißner-Institut, Bayerische Akademie der Wissenschaften, 85748 Garching, Germany}

\author{Dennis Huber}
\affiliation{\footnotesize Technische Universität München, Department Chemie, Lichtenbergstrasse 4, 85747 Garching, Germany}
\affiliation{\footnotesize Munich Center for Quantum Science and Technology (MCQST), 80799 München, Germany}
\author{Steffen J. Glaser}
\email{glaser@tum.de}
\affiliation{\footnotesize Technische Universität München, Department Chemie, Lichtenbergstrasse 4, 85747 Garching, Germany}
\affiliation{\footnotesize Munich Center for Quantum Science and Technology (MCQST), 80799 München, Germany}
\begin{abstract}
We present an experimental scanning-based tomography approach for near-term quantum devices. The underlying method has previously been introduced in an ensemble-based NMR setting. Here we provide a tutorial-style explanation along with suitable software tools to guide experimentalists in its adaptation to near-term pure-state quantum devices. The approach is based on a Wigner-type representation of quantum states and operators. These representations provide a rich visualization of quantum operators using shapes assembled from a linear combination of spherical harmonics. These shapes (called droplets in the following) can be experimentally tomographed by measuring the expectation values of rotated axial tensor operators. We present an experimental framework for implementing the scanning-based tomography technique for circuit-based quantum computers and showcase results from IBM quantum experience. We also present a method for estimating the density and process matrices from experimentally tomographed Wigner functions (droplets). This tomography approach can be directly implemented using the Python-based software package \texttt{DROPStomo}.           
\end{abstract}
\maketitle
\section{Introduction}
\label{intro}
Quantum tomography is an essential tool in quantum information processing to characterize quantum systems. Here, we use a phase-space tomography approach to recover finite-dimensional Wigner representations, with a particular focus on the DROPS (Discrete Representation of OPeratorS) representation~\cite{DROPS_main}. The DROPS representation follows the general strategy of Stratonovich~\cite{Stratonovich}, which specifies criteria for the definition of continuous Wigner functions for finite-dimensional quantum systems. This representation is based on the mapping of an arbitrary operator to a set of spherical functions denoted as \textit{droplets}. It provides an intuitive visualization approach to better understand quantum systems and quantum dynamics. An example of visualization of a two-qubit state is shown in Fig.~\ref{fig:intro_drops} and a detailed summary of the visualization approach is provided in supplementary Sec.~\ref{Supp:visualization}. This interactive DROPS visualization is implemented in the free SpinDrops~\cite{spinDROPS} software.
\\

The characteristic shapes of the \textit{droplets} arising from the DROPS representation is an outcome of an abstract mapping. In our recent studies~\cite{leiner2017wigner,leiner2018wigner}, these shapes were related to experimentally measurable quantities, and a scanning-based tomography approach was developed to measure the droplets corresponding to quantum states and unitary processes. The procedures were experimentally implemented on an NMR quantum information processor~\cite{Cory1634}, an ensemble quantum computer, where expectation values of observables can be directly measured~\cite{JONES201191}. This paper adapts the formalism of scanning-based tomography and presents an approach to experimentally implement state and process tomography on a pure-state quantum computer. In contrast to an NMR quantum information processor, on a pure-state quantum computer, expectation values are measured by many repetitions of projective measurements on individual quantum systems~\cite{nielsen2002quantum}. This study shows the flexibility of the scanning-based Wigner tomography approach and provides a tutorial-style approach for its implementation on current state-of-the-art pure-state quantum computing devices. \\
\begin{figure*}[t]
	\centering
	\includegraphics[scale=0.85]{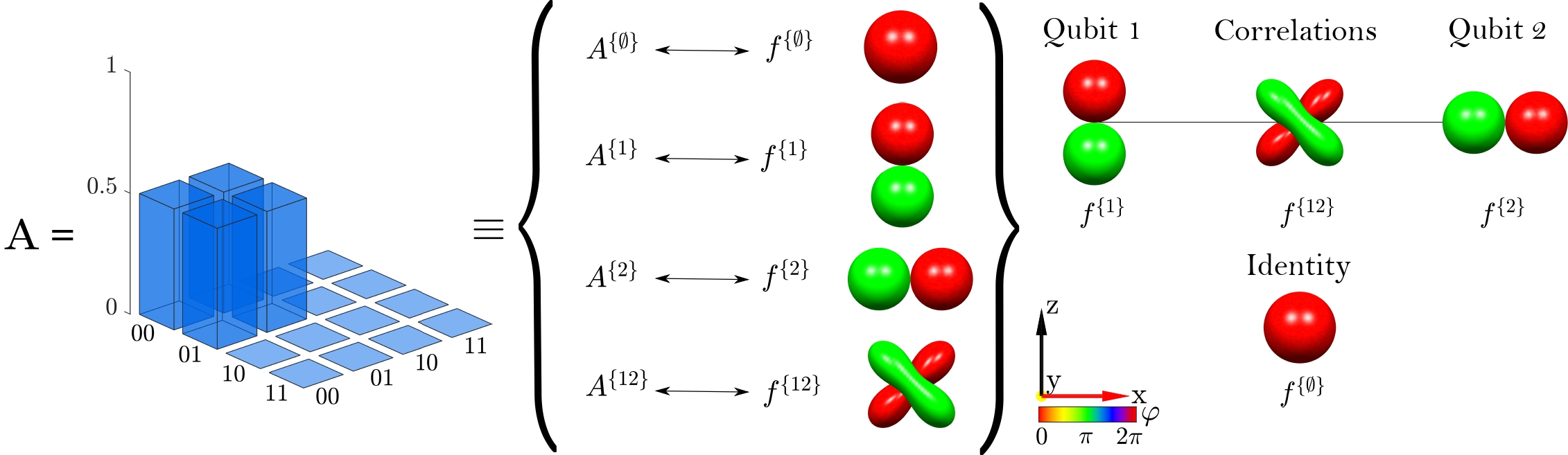}
	\caption{Skyscraper (left) and DROPS (right) visualization of a two-qubit quantum state $|\psi\rangle=\frac{1}{\sqrt{2}}(|00\rangle+|01\rangle)$ for which the density operator is given by the operator $A=\frac{1}{4}(\mathds{1}+\sigma_{1z}+\sigma_{2x}+\sigma_{1z}\sigma_{2x})$. The individual components $A^{\{\ell\}}$ of operator $A$ with label $\ell$ are mapped to spherical functions $f^{\{\ell\}}$ using a bijective mapping: $A = \sum_{\ell\in L} A^{(\ell)} \longleftrightarrow  \bigcup_{\ell\in L} f^{(\ell)}$. The droplets are combined in a systematic way in the rightmost panel which shows the droplets corresponding to state of first qubit ($\sigma_{1z} = \sigma_{z}\otimes\mathds{1}$), second qubit ($\sigma_{2x} = \mathds{1}\otimes\sigma_{x}$), correlations ($\sigma_{1z}\sigma_{2x} = \sigma_{z}\otimes\sigma_{x}$), and identity ($\mathds{1}$) terms. In these three-dimensional polar plot of droplets $f^{(\ell)}(\beta,\alpha)$, the distance from origin to a point on the surface is the absolute value $|f^{(\ell)}(\beta,\alpha)|$ and the color represents the phase $\varphi = \text{arg}[f^{(\ell)}(\beta,\alpha)]$ as defined by the color bar.} 
	\label{fig:intro_drops}
\end{figure*}

This particular tomography technique directly provides visual three-dimensional droplets based on experimental data, which helps to identify different experimental errors, such as gate imperfections, etc. We discuss this in more detail in Sec.~\ref{Sec.:DROPS_errors}, showcasing the physical intuitiveness of the DROPS representation in comparison with skyscraper visualizations~\cite{nielsen2002quantum}. We also illustrate how to estimate the matrix representation of density and process operators based on the experimentally measured droplet functions. The theory and experimental approaches presented here can be applied on any near-term quantum device. Here, we focus on the superconducting qubit-based IBM quantum platform for performing experiments. We also provide the Python-based software package \texttt{DROPStomo}~\cite{DROPStomo} related to this work. \texttt{DROPStomo} allows the use of Wigner state and process tomography on a simulator as well as on a quantum computer. This software package is discussed in Sec.~\ref{Sec.:Code}.            
\section{Illustration of main results}  
\label{main results}
\begin{figure}[h]
	\centering
	\includegraphics[scale=0.94]{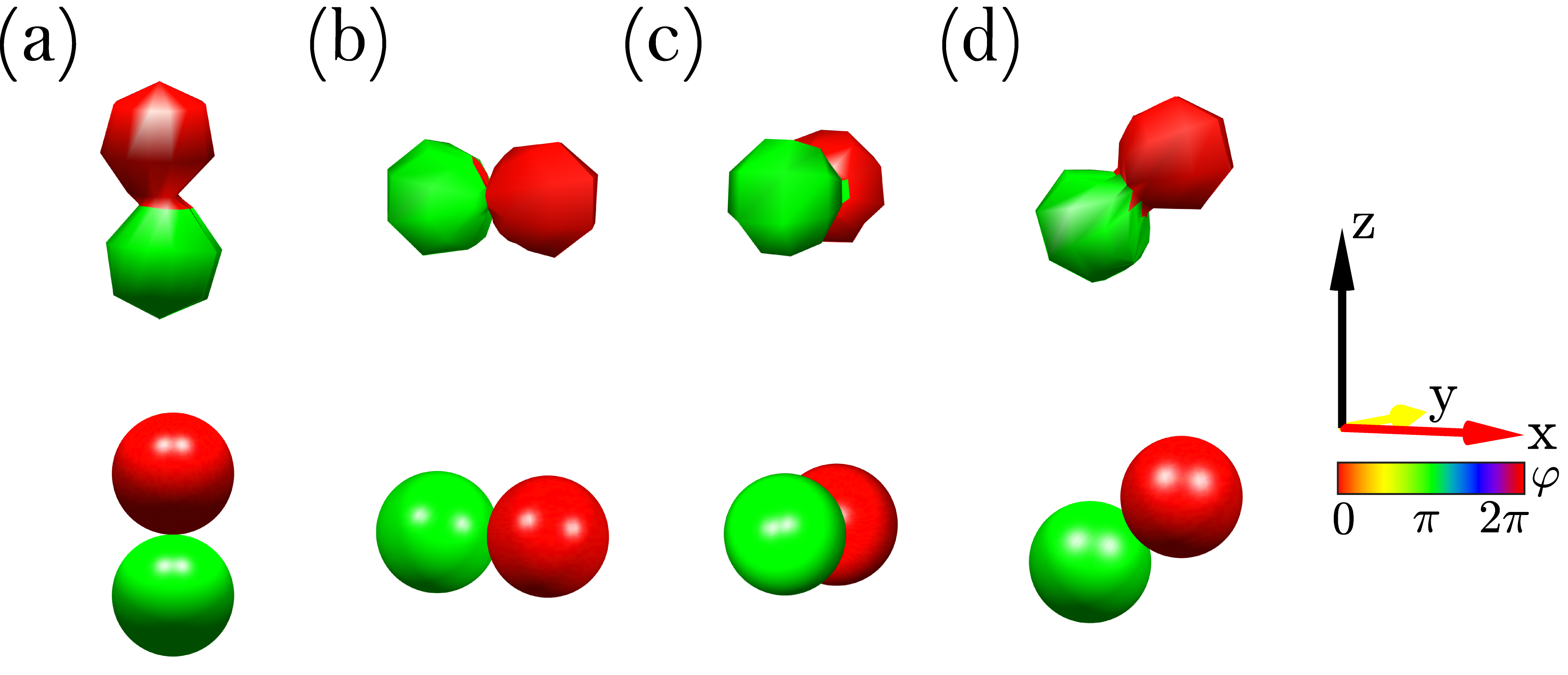}
	\caption{Experimentally tomographed (top panel) and simulated (lower panel) droplets corresponding to quantum states: (a) $|{0}\rangle$, (b) $\frac{|{0}\rangle+|{1}\rangle}{\sqrt{2}}$, (c) $\frac{|{0}\rangle+i|{1}\rangle}{\sqrt{2}}$, and (d) ${0.885|{0}\rangle+0.466|{1}\rangle}$.}
	\label{fig:QST result single qubit}
\end{figure}

\begin{figure}[h]
	\centering
	\includegraphics[scale=0.7]{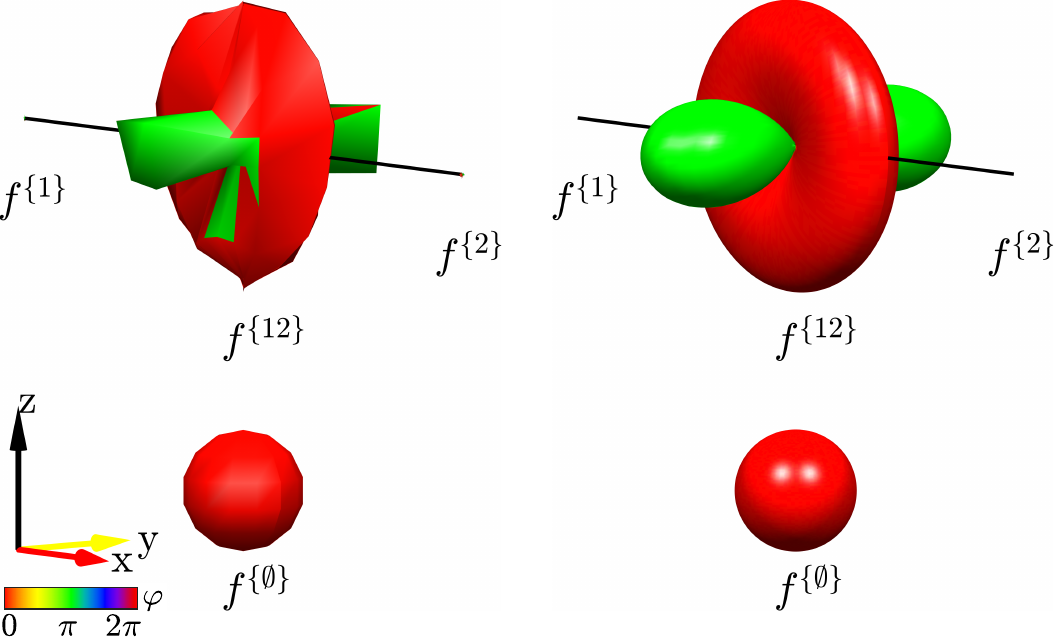}
	\caption{Experimentally tomographed (left) and simulated (right) DROPS representation of the Bell state $|\Phi^{+}\rangle = \frac{1}{\sqrt{2}}(|00\rangle+|11\rangle)$.}
	\label{fig:BellNew}
\end{figure}

\begin{figure}[h]
	\centering
	\includegraphics[scale=1.8]{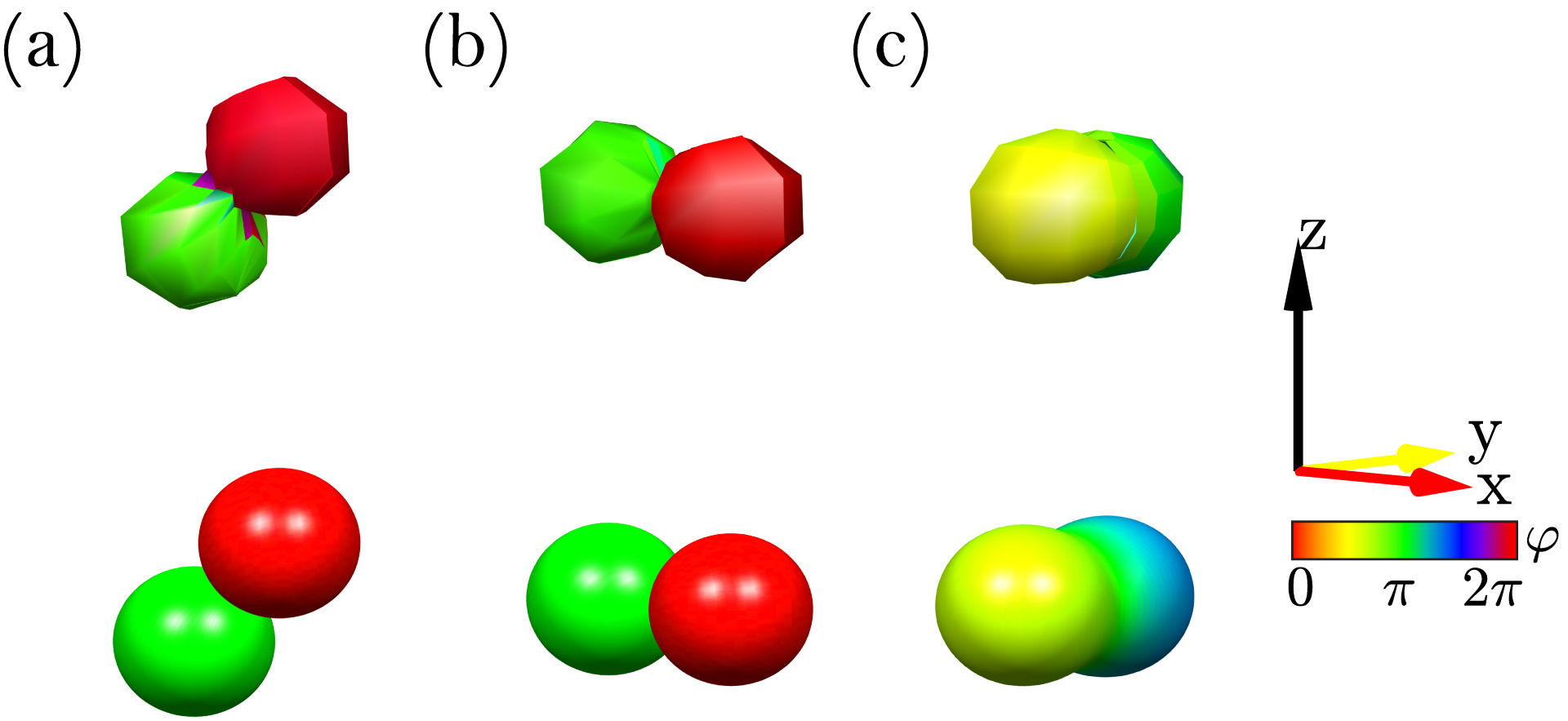}
	\caption{Experimentally tomographed (top panel), and simulated (lower panel) droplets for different quantum processes: (a) Hadamard gate, (b) NOT gate, and (c) $\big[\frac{3\pi}{2}\big]_{y}$ rotation.}
	\label{fig:QPT_results}
\end{figure}

In this section, we highlight the main results of our study. The scanning-based tomography approach estimates the expectation values of observable operators at different well-defined points on the Bloch sphere and combines them to form three-dimensional droplet functions. For example, in our experimental demonstration of state and process tomography, we use a simple equiangular sampling scheme with a combination of eight polar $\beta\in\{0,\frac{\pi}{7},\cdots\pi\}$ and fifteen azimuthal angles $\alpha\in\{0,\frac{2\pi}{14},\cdots2\pi\}$. This corresponds to a number of sample points $N_{p} = 8\cdot15 = 120$. More sophisticated sampling schemes are discussed in Sec.~\ref{Sec.:sampling_techqniues}. Experimental results for state tomography are shown in Fig.~\ref{fig:QST result single qubit} for a single-qubit system and in Fig.~\ref{fig:BellNew} for a two-qubit Bell state. 
For process tomography, the mapping of a unitary process matrix (U) onto a density matrix is a key step. This can be achieved by using an ancilla qubit and implementing a controlled process operation (cU) using the ancilla as a control qubit. Examples of process tomography results are showcased in Fig.~\ref{fig:QPT_results}. All the experiments in this study were performed on ibm$\_$lagos device with $N_{s} = 8192$ shots per sample point, i.e., for every combination of angles $\beta$, and $\alpha$. Overall, a total number of shots $N_{tot} = N_{s}\cdot N_{p}$ were acquired. The simulated droplets shown in the results are plotted with high resolution, whereas the experimental droplets are interpolated between the experimentally determined sampling points using the Matlab \textit{surf} function~\cite{MATLAB}. However, in the supplementary Sec.~\ref{Sec.:Add_result_figs}, we also show plots of the experimental and simulated droplets with the same resolution, both of which show comparable plotting artifacts of the Matlab display function.  

\section{Theory of Wigner quantum state tomography}
\label{QST_NISQ}
We are interested in experimentally scanning the Wigner representations of a density operator. This is a special case of quantum state tomography (QST), a vital tool in quantum computing and quantum information processing. Since the beginning of the field, there has been a lot of work performed in this direction~\cite{vogel1989determination,leonhardt1995quantum,white1999nonmaximally}. Recent studies use neural networks~\cite{torlai2018neural,neugebauer2020neural} and compressed sensing~\cite{gross2010quantum} to access the information about an unknown experimental state. Here we use a phase space~\cite{koczor2020continuous} tomography approach, which is helpful in experimentally visualizing quantum operators in finite-dimensional quantum systems. These, in general, can be any quantum operators such as density operators, quantum processes (propagators), etc. This section describes the scanning-based tomography approach for these operators.      
\subsection{Summary of scanning tomography approach}
\label{Scanning approach}
A general procedure for performing tomography in the context of Wigner representations using a scanning approach is described in the study~\cite{leiner2017wigner} (see results 1 and 2). Here we summarize this approach. In the following, we focus without loss of generality on a system consisting of $N$ qubits. Consider a multi-qubit quantum operator $A$, which is also represented by a set of rank $j$ and label $\ell$ spherical droplet functions $f^{(\ell)}=\sum_{j \in J(\ell)} f^{(\ell)}_{j} (\beta,\alpha)$ as described in Sec.~\ref{Supp:visualization}. To distinguish the size of different matrices, in the following, we refer to operators as $A^{[N]}$, where $N$ is the number of qubits and therefore, the operator $A^{[N]}$ is represented by a matrix of size $2^N\times2^N$. The main aim is to experimentally measure spherical droplet functions $f^{(\ell)}_{j}$ representing a quantum operator $A$ and this can be done by experimentally estimating the scalar products of rotated axial tensor operators $T_{j,\alpha\beta}^{(\ell)[N]}$ with operator $A^{[N]}$, where 
\begin{equation}
	\label{eq.7}
	T_{j,\alpha\beta}^{(\ell)[N]} = R_{\alpha\beta}^{[N]}(T_{j0}^{(\ell)})^{[N]}(R_{\alpha\beta}^{[N]})^\dagger.
\end{equation}
The term $(T_{j,\alpha\beta}^{(\ell)})^{[N]}$ is the rotated version of axial tensor operators $(T_{j0}^{(\ell)})^{[N]}$ of rank $j$ and order $m=0$. The rotation operator is given by  
\begin{equation}
	\label{eq.8}
	R_{\alpha\beta}^{[N]} = \text{exp}(-i\alpha F_{z}^{[N]})\text{exp}(-i\beta F_{y}^{[N]})
\end{equation}
where $F_{z} = \frac{1}{2}\sum_{k=1}^{N}\sigma_{kz}^{[N]}$, and $F_{y} = \frac{1}{2}\sum_{k=1}^{N}\sigma_{ky}^{[N]}$. $R_{\alpha\beta}^{[N]}$ corresponds to a rotation around the $y$ axis by polar angle $\beta\in[0,\pi]$ followed by rotation around the $z$ axis by azimuthal angle $\alpha\in[0,2\pi)$. Here we use the shorthand notation $\sigma_{ka} = \mathds{1}\otimes\dots\otimes\mathds{1}\otimes\sigma_{a}\otimes\mathds{1}\otimes\dots\otimes\mathds{1}$, where $\sigma_{a}$ is located on the $k^{th}$ position and $a\in\{x,y,z\}$. For given angles $\beta$ and $\alpha$, droplet function $f_{j}^{(\ell)}$ can be calculated by
\begin{equation}
	\label{eq.9a}
	f_{j}^{(\ell)}(\beta,\alpha) = s_{j}\langle{T_{j,\alpha\beta}^{(\ell)[N]}}|{A^{[N]}}\rangle,
\end{equation} 
which can be equivalently written using a shorthand notation as
\begin{equation}
	\label{eq.9b}
	f_{j}^{(\ell)}(\beta,\alpha) =  s_{j}\langle{T_{j,\alpha\beta}^{(\ell)[N]}}\rangle_{A^{[N]}},
\end{equation}        
where $s_{j} = \sqrt{(2j+1)/(4\pi)}$ and the scalar product $\langle{T_{j,\alpha\beta}^{(\ell)[N]}}|{A^{[N]}}\rangle$ is expressed as the expectation value of $T_{j,\alpha\beta}^{(\ell)[N]}$ for the density operator ${A^{[N]}}$:
\begin{equation}
	\label{eq.10}
	\langle{T_{j,\alpha\beta}^{(\ell)[N]}}\rangle_{A^{[N]}} = \text{tr}\big\{(T_{j,\alpha\beta}^{(\ell)[N]})A^{[N]}\big\}.
\end{equation}
Note that axial tensor operators are Hermitian, i.e., $({T_{j,\alpha\beta}^{(\ell)[N]}})^\dagger = ({T_{j,\alpha\beta}^{(\ell)[N]}})$~\cite{leiner2017wigner}.\\
\begin{figure}[ht]
	\centering
	\includegraphics[scale=0.9]{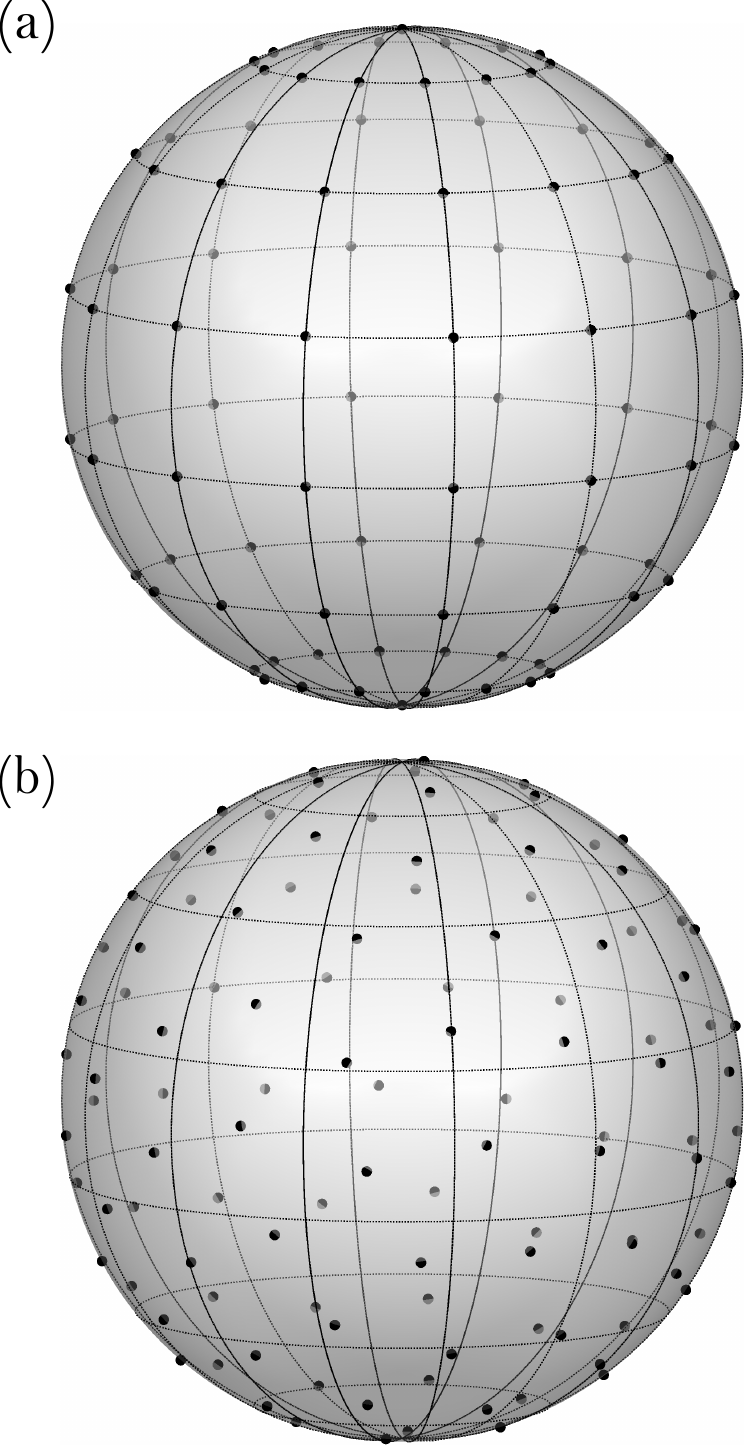}
	\caption{Two examples of different sampling schemes on a sphere for scanning: (a) equiangular and (b) REPULSION~\cite{BAK1997132}. The sampling points are plotted as dots on the sphere (faded dots represent sampling points located at the rear side of the sphere). Expectation values are computed for specified points on the sphere to tomograph the spherical functions representing a quantum operator $A$, as described in Eq.~\ref{eq.9b}.} 
	\label{fig:grid_plot}
\end{figure}

For the experimental tomography of droplet functions $f_{j}^{(\ell)}(\beta,\alpha)$ using the scanning approach, a multitude of choices for the set of sampling angles $\beta$ and $\alpha$ can be used. Fig.~\ref{fig:grid_plot} illustrates two such sampling techniques: equiangular and REPULSION~\cite{BAK1997132}. For simplicity, in the demonstration experiments shown here, we use an equiangular grid, but a numerical study using more sophisticated sampling techniques is presented and discussed in Sec.~\ref{Sec.:sampling_techqniues}.    	 
\subsection{Wigner quantum state tomography}
\label{Wigner_QST}
\begin{figure}[ht]
	\centering
	\includegraphics[scale=0.88]{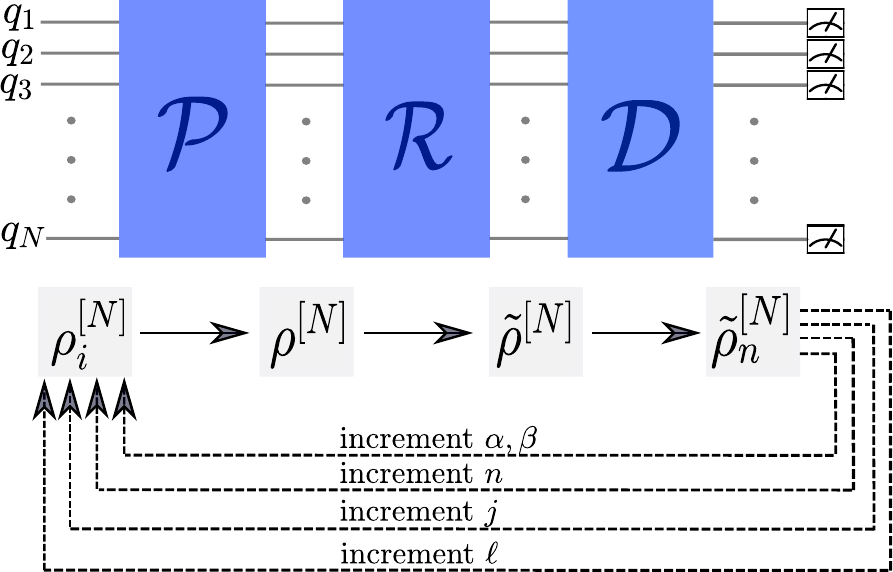}
	\caption{Schematic for the Wigner quantum state tomography algorithm. In general, the algorithm consists of three key blocks, namely Preparation ($\mathcal{P}$), Rotation ($\mathcal{R}$), and Detection-associated rotations ($\mathcal{D}$), which act on qubits $q_1, q_2,\dots,q_N$ and are followed by projective measurements. The lower part of the figure shows the evolution of the density matrix after each block. The algorithm is repeated for all desired combinations of parameters.}
	\label{fig:QST_algo}
\end{figure}

In this section, we first present an algorithm for Wigner quantum state tomography and then elaborate each step individually. A general step-wise procedure (see Fig.~\ref{fig:QST_algo}) to experimentally measure a spherical droplet function $f^{(\ell)}_{j}$ representing a density matrix $\rho^{[N]} = |\psi^{[N]}\rangle\langle\psi^{[N]}|$ is the following:
\begin{enumerate}[label=(\roman*)]
	\item \textbf{Preparation ($\mathcal{P}$)}: Prepare the desired quantum state $\rho^{[N]}$ from a defined initial state $\rho_{i}^{[N]}$. 
	\item \textbf{Rotation ($\mathcal{R}$)}: Rotate the density operator $\rho^{[N]}$ inversely for scanning.
	\item \textbf{Detection-associated rotations ($\mathcal{D}$)}: Apply local unitary operations to measure expectation values of Pauli operator components of axial tensor operators $T_{j0}^{(\ell)[N]}$ (see Table~\ref{Tab:AxialTensors}) that are not directly measurable.  
\end{enumerate}
These steps are repeated for the set of angles $\beta\in[0,\pi]$, $\alpha\in[0,2\pi]$ and for different local unitary operators $u_{n}$ (\textit{vide infra}), rank $j$, and label $\ell$ to experimentally scan the droplet functions $f_j^{(\ell)}$. In the rest of this section, we elaborate on each step of the presented algorithm.
\\

\textit{Step 1}: The first step of the algorithm, `Preparation', can be achieved by applying unitary operations depending on the initial and desired state. \\

\textit{Step 2}: Since our operator of interest is a density matrix $\rho^{[N]}$, Eq.~\ref{eq.9b} takes the following form:
\begin{equation}
	\label{eq.11}
	f_{j}^{(\ell)}(\beta,\alpha) = s_{j}\langle{T_{j,\alpha\beta}^{(\ell)[N]}}\rangle_{\rho^{[N]}}.
\end{equation} 
Instead of rotating the axial tensor operators $T_{j0}^{(\ell)[N]}$ as shown in Eq.~\ref{eq.7}, it is equivalent (and experimentally more convenient) to rotate the density matrix $\rho^{[N]}$ inversely, such that:
\begin{equation}
	\label{eq.12}
	f_{j}^{(\ell)}(\beta,\alpha) = s_{j}\langle{T_{j0}^{(\ell)[N]}}\rangle_{\tilde{\rho}^{[N]}},
\end{equation}
where
\begin{equation}
	\label{eq.13}
	\tilde{\rho}^{[N]} = 	(R_{\alpha\beta}^{[N]})^{-1}\rho^{[N]}R_{\alpha\beta}^{[N]}.
\end{equation}
The axial tensor operators $(T_{j0}^{(\ell)})$ are explicitly given in \cite{DROPS_main,leiner2020symmetry} for systems consisting of up to six qubits. In Table~\ref{Tab:AxialTensors}, we summarize the axial tensor operators for one and two-qubit systems.

\begin{table}[b]
	\caption{Axial tensor operators $T_{j0}^{(\ell)}$ for one ($N=1$) and two ($N=2$) qubit systems.}
	\centering
	\begin{tabular}{p{0.1cm} p{0.2cm} p{0.10cm} p{6cm}}
		\hline
		\hline
		$N$&\textrm{$\ell$}&\textrm{$j$}&\textrm{$T_{j0}^{(\ell)}$}\\[0.8ex]
		\hline
		1 & $\emptyset$ & 0 & $T_{00}^{(\emptyset)} = \frac{1}{\sqrt{2}}(\mathds{1})$\\ [0.7ex]
		
		& 1 & 1 & $T_{10}^{(1)} = \frac{1}{\sqrt{2}}(\sigma_{z})$\\[0.7ex]
		
		2 & $\emptyset$ & 0 & $T_{00}^{(\emptyset)} = \frac{1}{2}\mathds{1}$\\[0.7ex]
		
		& 1 & 1 & $T_{10}^{(1)} = \frac{1}{2}(\sigma_{1z})$\\[0.7ex]
		
		& 2 & 1 & $T_{10}^{(2)} = \frac{1}{2}(\sigma_{2z})$\\[0.7ex]
		
		& 12 & 0 & $T_{00}^{(12)} = \frac{1}{2\sqrt{3}}(\sigma_{1x}\sigma_{2x}+\sigma_{1y}\sigma_{2y}+\sigma_{1z}\sigma_{2z})$\\[0.7ex]
		
		& 12 & 1 & $T_{10}^{(12)}  = \frac{1}{2\sqrt{2}}(\sigma_{1x}\sigma_{2y}-\sigma_{1y}\sigma_{2x})$\\[0.7ex]
		
		& 12 & 2 & $T_{20}^{(12)} =\frac{-1}{2\sqrt{6}}(\sigma_{1x}\sigma_{2x}+\sigma_{1y}\sigma_{2y}-2\sigma_{1z}\sigma_{2z})$\\
		\hline
		\hline
	\end{tabular}
	\label{Tab:AxialTensors}
\end{table}

\textit{Step 3}: Depending on the number of qubits $N$ and rank $j$, the axial tensors $T_{j0}^{(\ell)}$ consist of different Pauli operators (see Table~\ref{Tab:AxialTensors}), but it might not be possible to measure these components directly depending on the specific quantum computing hardware. In a typical pure-state quantum computing device, the measurement is done along the $z$ axis, which implies that the directly measurable operators are: $\mathds{1}$, $\sigma_{1z}$, $\sigma_{2z}$, and $\sigma_{1z}\sigma_{2z}$. In this case, measuring expectation values of non-directly measurable operators can be achieved with the help of local unitary operations $u_{n}$. For example, consider the expectation value $\langle T_{10}^{(12)}\rangle_{\tilde{\rho}(\beta,\alpha)}$ given by:
\begin{equation}
	\label{eq.14}
	\small
	\begin{aligned}
		\langle T_{10}^{(12)}\rangle_{\tilde{\rho}(\beta,\alpha)} & = \frac{1}{2\sqrt{2}}\langle{\sigma_{1x}\sigma_{2y}-\sigma_{1y}\sigma_{2x}\rangle}_{\tilde{\rho}(\beta,\alpha)}\\
		& = \frac{1}{2\sqrt{2}}\langle{\sigma_{1x}\sigma_{2y}}\rangle_{\tilde{\rho}(\beta,\alpha)}-
		\frac{1}{2\sqrt{2}}\langle{\sigma_{1y}\sigma_{2x}}\rangle_{\tilde{\rho}(\beta,\alpha)}.
	\end{aligned}
\end{equation}
In the first term ($n=1$) of Eq.~\ref{eq.14}, the expectation value $\langle{\sigma_{1x}\sigma_{2y}}\rangle_{\tilde{\rho}(\beta,\alpha)}$ needs to be determined. This can be achieved by measuring instead the expectation value
\begin{equation}
	\langle{\sigma_{1z}\sigma_{2z}}\rangle_{\tilde{\rho}_{1}(\beta,\alpha)} = \langle{\sigma_{1x}\sigma_{2y}}\rangle_{\tilde{\rho}(\beta,\alpha)},
\end{equation}
where  
\begin{equation}
	\label{eq.16}
	\tilde{\rho}_{1}(\beta,\alpha) = u_{1} \tilde{\rho}(\beta,\alpha) u_{1}^{\dagger}.
\end{equation}
The density operator $\tilde{\rho}_{1}(\beta,\alpha)$ is obtained from $\tilde{\rho}(\beta,\alpha)$ by applying a $-\pi/2$ rotation around the $y$ axis (for bringing the $x$ axis to the $z$ axis) to the first qubit and a $\pi/2$ rotation around the $x$ axis (for bringing the $y$ axis to the $z$ axis) to the second qubit. This corresponds to the local unitary transformation $u_{1} = \big((R_{0,\frac{\pi}{2}})^{-1}\otimes\mathds{1}\big)\cdot(\mathds{1}\otimes (R_{\frac{\pi}{2},\frac{\pi}{2}})^{-1})$.\\ 

Similarly, in the second term ($n=2$) of Eq.~\ref{eq.14}, the expectation value $\langle{\sigma_{1y}\sigma_{2x}}\rangle_{\tilde{\rho}(\beta,\alpha)}$ needs to be determined. This can be achieved by measuring instead the expectation value
\begin{equation}
	\langle{\sigma_{1z}\sigma_{2z}}\rangle_{\tilde{\rho}_{2}(\beta,\alpha)} = \langle{\sigma_{1y}\sigma_{2x}}\rangle_{\tilde{\rho}(\beta,\alpha)},
\end{equation}
where  
\begin{equation}
	\label{eq.16}
	\tilde{\rho}_{2}(\beta,\alpha) = u_{2} \tilde{\rho}(\beta,\alpha) u_{2}^{\dagger}.
\end{equation}
The density operator $\tilde{\rho}_{2}(\beta,\alpha)$ is obtained from $\tilde{\rho}(\beta,\alpha)$ by applying a $\pi/2$ rotation around the $x$ axis (for bringing the $y$ axis to the $z$ axis) to the first qubit and a $-\pi/2$ rotation around the $y$ axis (for bringing the $x$ axis to the $z$ axis) to the second qubit. This corresponds to the local unitary transformation $u_{2} = \big((R_{{\frac{\pi}{2}},\frac{\pi}{2}})^{-1}\otimes\mathds{1}\big)\cdot(\mathds{1}\otimes (R_{0,\frac{\pi}{2}})^{-1})$. \\

Overall, for $\ell=\{12\}$ and $j=1$, the droplet function from Eq.~\ref{eq.12} can be expressed as
\begin{equation}
	\begin{aligned}
		f_{1}^{(12)}(\beta,\alpha) & = s_{1}\langle T_{10}^{(12)}\rangle_{\tilde{\rho}(\beta,\alpha)} \\& =\frac{s_{1}}{2\sqrt{2}}(\langle{\sigma_{1z}\sigma_{2z}}\rangle_{\tilde{\rho}_{1}(\beta,\alpha)}+\langle{\sigma_{1z}\sigma_{2z}}\rangle_{\tilde{\rho}_{2}(\beta,\alpha)}).
	\end{aligned}
	\label{eq.15}
\end{equation}  
Hence, only projective measurements along the $z$ axis are required, c.f.~\ref{appendix A}. 
\subsection{Estimation of a density matrix from the droplet functions}
\label{subsec.:estimation}
In this section, we show how the \textit{matrix} form of a density operator can be estimated based on its experimentally measured DROPS representation. A general $N$-qubit density matrix~\cite{nielsen2002quantum} can be expressed in terms of Pauli operators as:   
\begin{equation}
	\label{eq.17}
	\rho^{[N]} = \sum_{a=0}^3\sum_{b=0}^3\dots\sum_{g=0}^3 r_{ab\dots g} 	(\sigma_{a}\otimes\sigma_{b}\otimes\dots\otimes\sigma_{g})
\end{equation} 
where $\sigma_{0}$ is $\mathds{1}$ (the 2$\times$2 identity matrix), while $\sigma_{1},\sigma_{2}$, and $\sigma_{3}$ are standard Pauli matrices $\sigma_{x},\sigma_{y}$, and $\sigma_{z}$ respectively. The terms $r_{ab\dots g}$ are real coefficients. Given the DROPS representation of a density operator, these coefficients can be computed by calculating the scalar product between basis droplets (ideally simulated without noise) and experimental droplets~\cite{leiner2017wigner}. The basis droplets can be generated using the definitions provided in supplementary Sec.~\ref{Sec.:Sph_basis} for one and two qubits.\\

The scalar product between two tensor operators can be approximated by the \textit{discretized} scalar product between their droplet functions, see supplementary Sec.~\ref{appendix B}. In the general case of a droplet $f_A$ with complex values $f_{A}(\theta_{i},\phi_{i})$ and another droplet $f_B$ with complex values $f_{B}(\theta_{i},\phi_{i})$ at the grid points, the scalar product or overlap between the two droplets is given by
\begin{equation}
	\label{eq.18}
	r = \langle f_{A}|f_{B}\rangle =  \sum_{i} \text{w}_{i} f_{A}^{*}(\theta_{i},\phi_{i})f_{B}(\theta_{i},\phi_{i}),
\end{equation}
where $f_{A}^{*}(\theta_{i},\phi_{i})$ is the complex conjugate of $f_{A}(\theta_{i},\phi_{i})$. The sampling weights $\text{w}_{i}$ corresponding to an equiangular grid for calculating the scalar product are provided in supplementary Sec.~\ref{appendix B}. The sampling weights for other sampling techniques, such as REPULSION and Lebedev, are available in~\cite{spinach}, and~\cite{Lebedev_Rule}, respectively. \\

Based on the estimated experimental density operator $\rho^{[N]}$, the state fidelity ($\mathcal{F}_s$)~\cite{liang2019quantum} with which a desired state $\rho_t^{[N]}$ has been reached can be calculated using the normalized scalar product:
\begin{equation}
	\label{eq.Sf}
	\mathcal{F}_s = \frac{\text{tr}(\rho^{[N]} \rho_t^{[N]})}{\sqrt{\text{tr}((\rho^{[N]})^2) \text{tr}((\rho_t^{[N]})^2)}}.
\end{equation}

In the next section, we focus on the experimental implementation of the presented algorithm for one and two-qubit systems and showcase the experimental results performed on the ibm\_lagos device.
\section{Experimental implementation of Wigner state tomography}
\label{sec.:QST experiment}
In this section we discuss how the Wigner quantum state tomography algorithm can be implemented on an experimental near-term quantum device. For concreteness, we will present the implementation using Qiskit \cite{gadi_aleksandrowicz_2019_2562111}, an open-source quantum development platform for simulations and experiments. Wigner state tomography can be used directly for one- and two-qubit systems using the Python-based software package \texttt{DROPStomo}~\cite{DROPStomo}.    
\subsection{One qubit}
\label{Sec.:one qubit QST}
For a system consisting of one qubit ($N=1$), there are only two possible values for the rank $j$: $j=0$ for $\ell=\{\emptyset\}$ and $j=1$ for $\ell=\{1\}$. Hence, the single-qubit density matrix $\rho^{[1]}$ represented by spherical functions $f_{0}^{(\emptyset)}$ and $f_{1}^{(1)}$ can be tomographed by measuring the expectation values from Eq.~\ref{eq.12} as
\begin{equation}
	\label{eq.19}
	\begin{aligned}
		f_{0}^{(\emptyset)}(\beta,\alpha) & = \sqrt{\dfrac{1}{4\pi}}\langle{(T_{00}^{(\emptyset)})^{[1]}}\rangle_{\tilde{{\rho}}^{[1]}},\\
		f_{1}^{(1)}(\beta,\alpha) & = \sqrt{\dfrac{3}{4\pi}}\langle{(T_{10}^{(1)})^{[1]}}\rangle_{\tilde{{\rho}}^{[1]}}.
	\end{aligned}
\end{equation}
Substituting the explicit form of the tensor operators $T_{00}^{(\emptyset)}$ and $T_{10}^{(1)}$ from Table~\ref{Tab:AxialTensors} gives 
\begin{equation}
	\label{eq.20}
	\begin{aligned}
		f_{0}^{(\emptyset)}(\beta,\alpha) &= \sqrt{\dfrac{1}{8\pi}}\langle{\mathds{1}}\rangle_{\tilde{{\rho}}^{[1]}},\\
		f_{1}^{(1)}(\beta,\alpha) &= \sqrt{\dfrac{3}{8\pi}}\langle{\sigma_{z}}\rangle_{\tilde{{\rho}}^{[1]}},
	\end{aligned}
\end{equation}
where $\mathds{1}$ is the $2\times2$ identity matrix. The required expectation values can be computed directly on a pure-state quantum computer based on the outcome probabilities $p_0$ and $p_1$ corresponding to state $|0\rangle$ and state $|1\rangle$ respectively:
\begin{equation}
	\label{eq.20a}
	\begin{aligned}
		\langle{\mathds{1}}\rangle& = p_{0}+p_{1}, \\
		\langle{\sigma_{z}}\rangle& = p_{0}-p_{1}.
	\end{aligned}
\end{equation}
The probabilities $p_{0}$ and $p_{1}$ can be experimentally measured by repeating an experiment multiple times, where the number of repetitions is also referred to as number of shots $N_{s}$. We discuss this in more detail in supplementary Sec.~\ref{appendix A}. In the considered case, the detection-associated rotation ($\mathcal{D}$) step is not required. \\

The probabilities $p_{0}$ and $p_{1}$ required for computing the expectation values for droplets $f_0^{(\emptyset)}$ and $f_1^{(1)}$ can be measured in the same experiment; hence the algorithm does not require to be repeated for rank $j$ and label $\ell$. In supplementary Sec.~\ref{Sec.:Qcircuits}, we explicitly provide the quantum circuit (Fig.~\ref{fig:QST1_1}) for performing tomography of a qubit in the state $|\psi\rangle = \frac{1}{\sqrt{2}}(|0\rangle+|1\rangle$ along with the corresponding plot of simulated and experimental expectation values (Fig.~\ref{fig:expec&drops}) of droplet function $f_1^{(1)}(\beta,\alpha)$.\\ 

Based on the experimentally tomographed droplet functions $f_{0}^{(\emptyset)}$ and $f_{1}^{(1)}$, the corresponding density matrix can be estimated as discussed in Sec.~\ref{subsec.:estimation}. For a single qubit, the density matrix can be expressed in terms of Pauli operators as
\begin{equation}
	\label{densityMat}
	\rho^{[1]} = 
	\sum_{k=0}^3 r_k \sigma_{k}. 
\end{equation}
To compute the coefficient $r_0$, a scalar product is calculated using Eq.~\ref{eq.18} between the ideal basis droplet ($f_{\sigma_{0}}$) and the experimentally tomographed $f_{0}^{(\emptyset)}$ droplet:
\begin{equation}
	\label{eq.24n}
	r_{0} = \langle f_{\sigma_{0}}|f_{0}^{(\emptyset)}\rangle.
\end{equation}
To compute the remaining coefficients $r_k$ for $k\in\{1,2,3\}$, 
we calculate the scalar product between all ideal basis droplets with label $\ell=1$, rank $j=1$, and the experimentally tomographed rank $j=1$ droplet $f_{1}^{(1)}$:
\begin{equation}
	\label{eq.24n1}
	r_{k} = \langle f_{\sigma_{k}}|f_{1}^{(1)}\rangle.
\end{equation}
The ideal basis droplet functions $f_{\sigma_{k}}$ are provided in supplementary Sec.~\ref{Sec.:Basis_single}. Hence, by calculating these overlap coefficients $r_k$, an estimate of the density matrix can be obtained using Eq.~\ref{densityMat}. In Table~\ref{table: single qubit fid}, we provide experimental state tomography fidelities computed using Eq.~\ref{eq.Sf} for different states considered in Fig.~\ref{fig:QST result single qubit}. The standard deviations given in Table~\ref{table: single qubit fid} were estimated by conducting the experiment three times with the state $|\psi_t\rangle = 0.885|0\rangle+0.466|1\rangle$, for a reference. To avoid redundancy, in Fig.~\ref{fig:QST result single qubit}, we only plot the rank $j=1$ droplets, as the rank $j=0$ droplet is a sphere of radius $\sqrt{1/(8\pi)}$ in the single-qubit case. 	
\begin{table}[h]
	\caption{
		Experimental state tomography fidelities ($\mathcal{F}_s$) corresponding to the desired single-qubit quantum states $|\psi_t\rangle$, see Fig.~\ref{fig:QST result single qubit}.}
	\centering
	\begin{tabular}{p{4.5cm} p{2.5cm}}
		\hline
		\hline
		$|\psi_t\rangle$&$\mathcal{F}_s$\\[0.6ex]
		\hline
		$\frac{1}{\sqrt{2}}(|0\rangle+|1\rangle)$   &0.9991$\pm1\times10^{-3}$\\[0.8ex]
		$|0\rangle$  & 0.9991$\pm1\times10^{-3}$ \\[0.8ex]
		$\frac{1}{\sqrt{2}}(|0\rangle+i|1\rangle)$   &0.9992$\pm1\times10^{-3}$\\[0.8ex]
		$0.885|0\rangle+0.466|1\rangle$   &0.9990$\pm1\times10^{-3}$\\
		\hline
		\hline
	\end{tabular}
	\label{table: single qubit fid}
\end{table} 

\subsection{Two qubits}
\label{Sec.:two qubit QST}
\begin{figure}[h]
	\centering
	\includegraphics[scale=0.77]{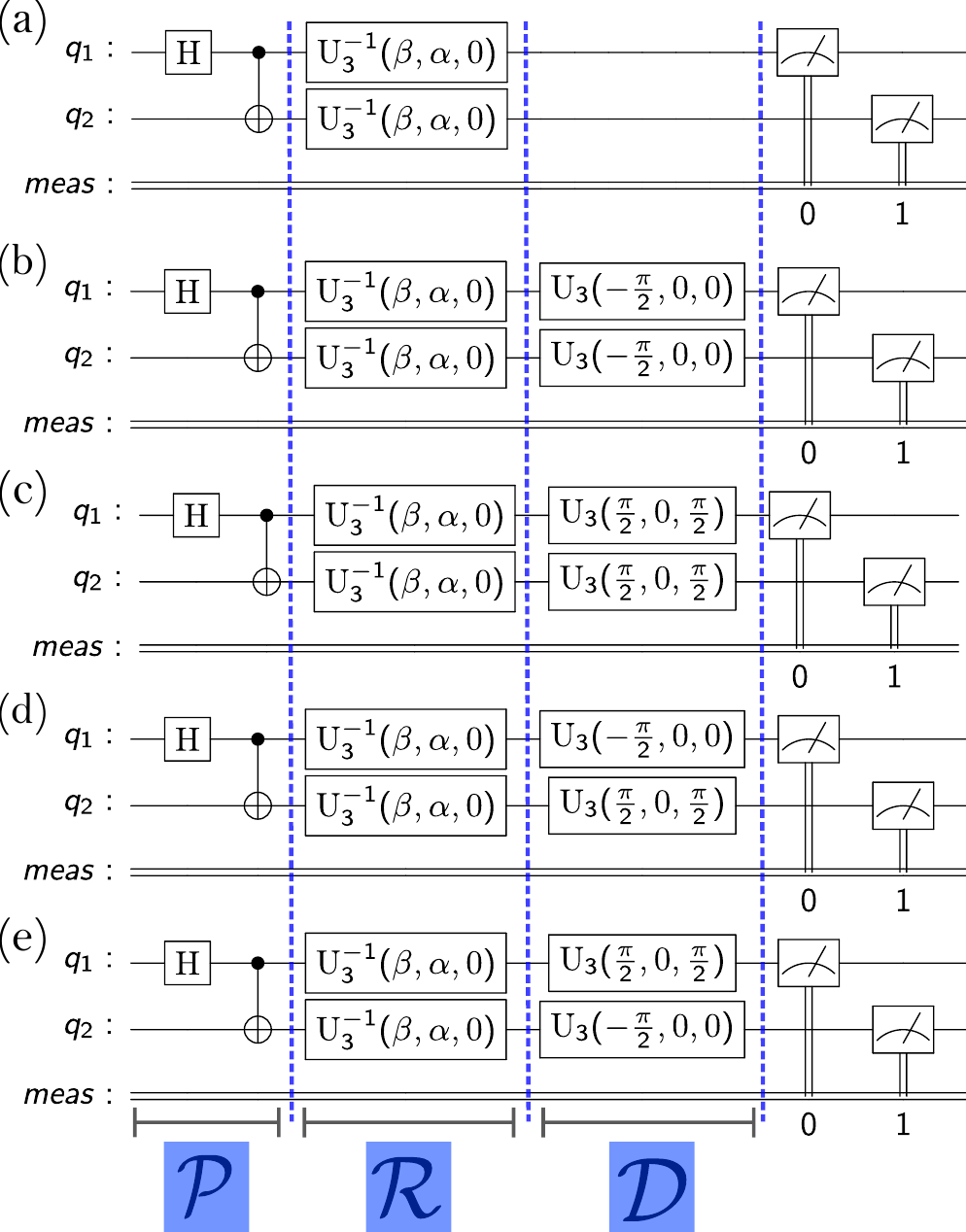}
	\caption{Quantum circuit set for a two-qubit Wigner state tomography of a Bell state $|\psi\rangle = \frac{1}{\sqrt{2}}(|00\rangle+|11\rangle)$. The Bell state is prepared from the initial state $|\psi\rangle_{i} = |00\rangle$ by applying the Hadamard (H) gate on $q_1$ followed by a controlled-NOT gate. The three blocks Preparation ($\mathcal{P}$), Rotation ($\mathcal{R}$), and Detection-associated rotations ($\mathcal{D}$) are shown here. Circuit (a) provides the expectation values for the operators $\mathds{1},\sigma_{1z},\sigma_{2z}$, and $\sigma_{1z}\sigma_{2z}$. Similarly, circuits (b), (c), (d), and (e) provide the expectation values for the operators $\sigma_{1x}\sigma_{2x},\sigma_{1y}\sigma_{2y},\sigma_{1x}\sigma_{2y}$, and $\sigma_{1y}\sigma_{2x}$, respectively. The $\mathrm{U}_{3}$ gate used in the circuit is discussed in the supplementary Sec.~\ref{supp:U3}.}
	\label{fig:QSTCircuit_two}
\end{figure}

For a two-qubit system ($N=2$), there are four possible labels $\ell$, and based on these labels, there are different ranks $j$ as shown in Table~\ref{Tab:AxialTensors}. Therefore, for a two-qubit system, the Wigner quantum state tomography requires measuring the following spherical droplets $f_{j}^{(\ell)}$:
\begin{equation}
	\label{eq.25}
	\begin{split}
		f_{0}^{(\emptyset)}(\beta,\alpha) & = \sqrt{\dfrac{1}{4\pi}}\expval*{(T_{00}^{(\emptyset)})^{[2]}}_{\tilde{{\rho}}^{[2]}}\\
		& =  \dfrac{1}{4\sqrt{\pi}}\expval*{\mathds{1}}_{{\tilde{\rho}}^{[2]}}
	\end{split}
\end{equation}
for the identity droplet ($j$=0, and $\ell$=$\emptyset$), here $\mathds{1}$ is the 4$\times$4 identity matrix. For droplets of rank $j=1$, for each qubit (with labels $\ell=1$ and $\ell=2$) we have:
\begin{equation}
	\label{eq.26}
	\begin{split}
		f_{1}^{(1)}(\beta,\alpha) & = \sqrt{\dfrac{3}{4\pi}}\expval*{(T_{10}^{(1)})^{[2]}}_{\tilde{{\rho}}^{[2]}}\\
		& =  \frac{1}{4}\sqrt{\dfrac{3}{\pi}}\expval*{\sigma_{1z}}_{\tilde{{\rho}}^{[2]}},
	\end{split}
\end{equation} 
\begin{equation}
	\label{eq.27}
	\begin{split}
		f_{1}^{(2)}(\beta,\alpha) & = \sqrt{\dfrac{3}{4\pi}}\expval*{(T_{10}^{(2)})^{[2]}}_{\tilde{{\rho}}^{[2]}}\\
		& =  \frac{1}{4}\sqrt{\dfrac{3}{\pi}}\expval*{\sigma_{2z}}_{\tilde{{\rho}}^{[2]}}.
	\end{split}
\end{equation} 
For bilinear terms with label $\ell=12$, we have
\begin{equation}
	\label{eq.28}
	\small
	\begin{split}
		f_{0}^{(12)}(\beta,\alpha) & = \sqrt{\dfrac{1}{4\pi}}\expval*{(T_{00}^{(12)})^{[2]}}_{\tilde{{\rho}}^{[2]}}\\
		& =  \dfrac{1}{4\sqrt{3\pi}}\expval*{(\sigma_{1x}\sigma_{2x}+\sigma_{1y}\sigma_{2y}+\sigma_{1z}\sigma_{2z})}_{\tilde{{\rho}}^{[2]}}
	\end{split}
\end{equation} 
for rank $j=0$, 
\begin{equation}
	\label{eq.29}
	\begin{split}
		f_{1}^{(12)}(\beta,\alpha) & = \sqrt{\dfrac{3}{4\pi}}\expval*{(T_{10}^{(12)})^{[2]}}_{\tilde{{\rho}}^{[2]}}\\
		& =  \dfrac{1}{4}\sqrt{\dfrac{3}{2\pi}}\expval*{(\sigma_{1x}\sigma_{2y}-\sigma_{1y}\sigma_{2x})}_{\tilde{{\rho}}^{[2]}}
	\end{split}
\end{equation}
for rank $j=1$, and
\begin{equation}
	\label{eq.30}
	\small
	\begin{split}
		f_{2}^{(12)}(\beta,\alpha) & = \sqrt{\dfrac{5}{4\pi}}\expval*{(T_{20}^{(12)})^{[2]}}_{\tilde{{\rho}}^{[2]}}\\
		& =  \dfrac{1}{4}\sqrt{\dfrac{5}{6\pi}}\expval*{(-\sigma_{1x}\sigma_{2x}-\sigma_{1y}\sigma_{2y}+2\sigma_{1z}\sigma_{2z})}_{\tilde{{\rho}}^{[2]}}
	\end{split}
\end{equation}
for rank $j=2$. Different Pauli operator expectation values are required in Eq.~\ref{eq.26} to Eq.~\ref{eq.30}, and some of them are not directly measurable. In this case, we use the last block of the algorithm called detection-associated rotations ($\mathcal{D}$) as explained with an example in Sec.~\ref{Wigner_QST}. The rotations required for step $\mathcal{D}$ can be implemented in terms of local $\mathrm{U}_3$ gates as described in supplementary Sec.~\ref{supp:U3}. Similar to the computation of expectation values of linear terms in Eq.~\ref{eq.20a}, the expectation values of bilinear terms can be computed by combining the outcome probabilities:  
\begin{equation}
	\label{eq.30a}
	\begin{aligned}
		\langle\mathds{1}\rangle &= 			p_{00}+p_{01}+p_{10}+p_{11},\\
		\langle\sigma_{1z}\sigma_{2z}\rangle &= 			p_{00}-p_{01}-p_{10}+p_{11},
	\end{aligned}	
\end{equation}
where $p_{ab}$ for $a,b\in\{0,1\}$ is the probability corresponding to state $|ab\rangle$. We refer to supplementary Sec.~\ref{appendix A} for more information. In Fig.~\ref{fig:QSTCircuit_two}, we explicitly show the quantum circuits for state tomography of a Bell state $|\psi\rangle = \frac{1}{\sqrt{2}}(|00\rangle+|11\rangle)$. The different circuits in the figure are used for calculating the different expectation values, which are then combined to form a particular droplet function $f_{j}^{(\ell)}$. Hence, for a two-qubit Wigner quantum state tomography, a maximum of five quantum circuits are required, which are repeated for all combinations of values of the angles $\beta$ and $\alpha$.  
 \\

\begin{figure}[h]
	\centering
	\includegraphics[scale=1]{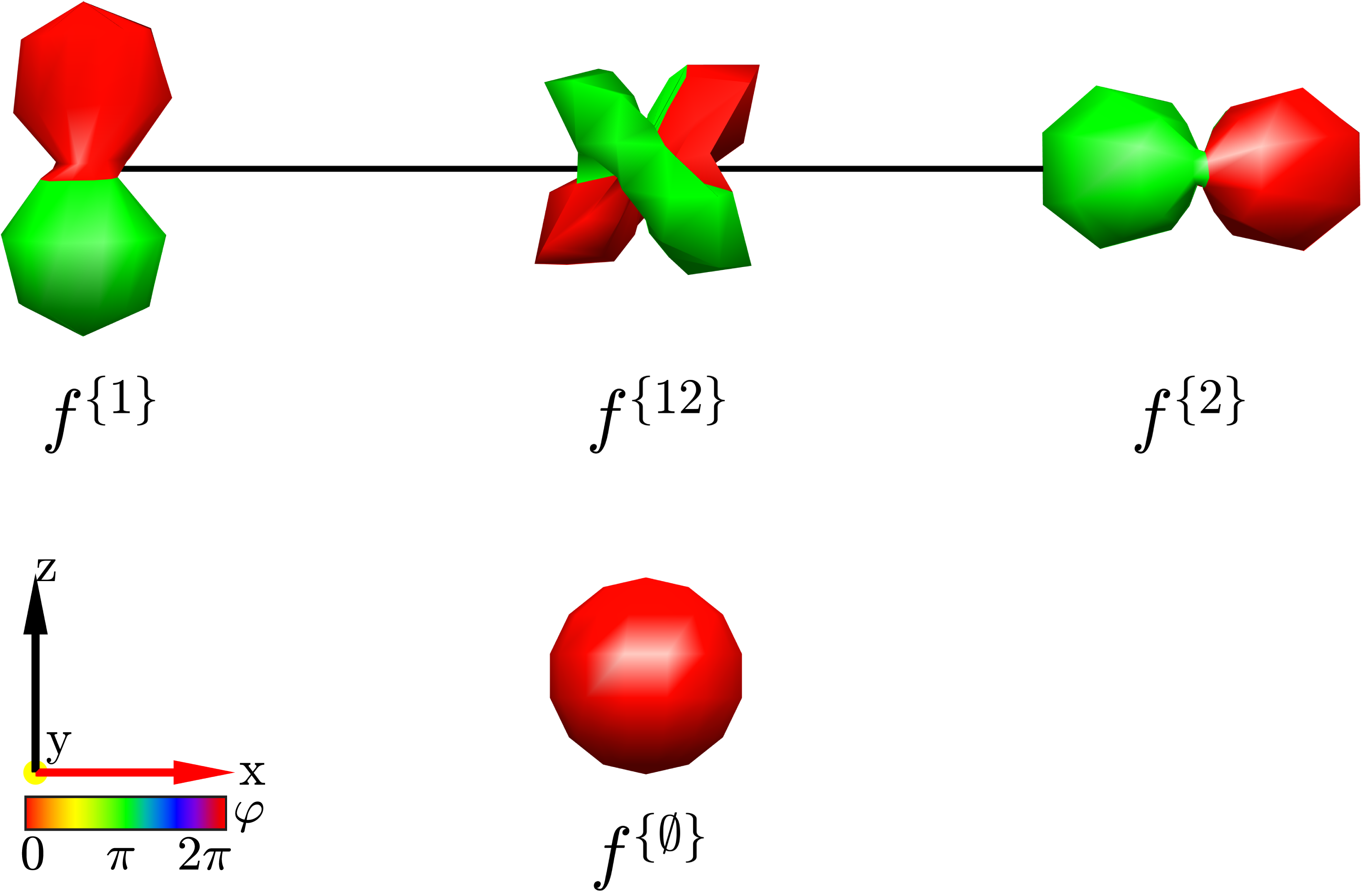}
	\caption{Experimentally tomographed DROPS representation of a two-qubit state $|\psi\rangle = \frac{1}{\sqrt{2}}(|00\rangle+|01\rangle)$. See Fig.~\ref{fig:intro_drops} for simulated droplets.}
	\label{fig:00+01New}
\end{figure}

Similar to the one-qubit system, we can estimate the density matrix based on the experimentally tomographed droplets. The density matrix for a two-qubit system can be expressed in terms of Pauli operators as
\begin{equation}
	\label{eq.30n}
	\rho^{[2]} = \sum_{k=0}^3\sum_{l=0}^3 r_{kl} (\sigma_{k}\otimes\sigma_{l}).
\end{equation}
The terms $r_{kl}$ with $k,l\in\{0,1,2,3\}$ are real coefficients and can be calculated by computing the scalar product between the droplet functions as shown in Eq.~\ref{eq.18}. To compute $r_{00}$, the scalar product is calculated between the simulated ideal basis droplet ($f_{\sigma_{0}}$) with label $\ell = \emptyset$, rank $j=0$ and the experimentally tomographed droplet function $f_{0}^{(\emptyset)}$ as, 
\begin{equation}
	\label{eq.30n1}
	r_{00} = \langle f_{\sigma_{0}}|f_{0}^{(\emptyset)}\rangle.
\end{equation}
To calculate the coefficients $r_{k0}$ for $k\in\{1,2,3\}$ the scalar product is computed between all ideal basis droplets with label $\ell=1$, rank $j=1$ and the experimentally tomographed droplet $f_{1}^{(1)}$: 
\begin{equation}
	\label{eq.30n2}
	r_{k0} = \langle f_{\sigma_{1k}}|f_{1}^{(1)}\rangle.
\end{equation}
Similarly, the coefficients $r_{0l}$ for $l\in\{1,2,3\}$ can be computed by calculating the scalar product between the ideal basis droplets of label $\ell=2$, rank $j=1$ and experimentally tomographed droplet $f_1^{(2)}$:
\begin{equation}
	\label{eq.30n3}
	r_{0l} = \langle f_{\sigma_{2l}}|f_{1}^{(2)}\rangle.
\end{equation}
The remaining bilinear coefficients $r_{kl}$ for $k,l\in\{1,2,3\}$ can be calculated by computing the scalar product between the ideal bilinear basis droplets $f_{\sigma_{1k}\sigma_{2l}}$ with the sum of the experimentally tomographed droplets $f_0^{(12)}$, $f_1^{(12)}$, and $f_2^{(12)}$:
\begin{equation}
	\label{eq.30n4}
	r_{kl} = \langle f_{\sigma_{1k}\sigma_{2l}}|f_{0}^{(12)}+f_{1}^{(12)}+f_{2}^{(12)}\rangle.
\end{equation}
The ideal basis droplets are provided in the supplementary Sec.~\ref{Sec.:Basis_two}. Hence, by calculating these coefficients for every value of $k$ and $l$, a density matrix can be estimated using Eq.~\ref{eq.30n} and the state fidelity ($\mathcal{F}_s$) can be computed using Eq.~\ref{eq.Sf}. In Table~\ref{table: two qubit fid}, we present the experimental state fidelities of two-qubit examples. The standard deviation given in Table~\ref{table: two qubit fid} were estimated by conducting the experiment three times with the state $|\psi_t\rangle = \frac{1}{\sqrt{2}}(|00\rangle+|11\rangle)$, for a reference. In Fig.~\ref{fig:BellNew} and Fig.~\ref{fig:00+01New}, we show the experimentally tomographed and theoretical droplets for a maximally entangled Bell state and a separable quantum state, respectively. The bilinear droplets are combined to $f^{(12)} = f_{0}^{(12)}+f_{1}^{(12)}+f_{2}^{(12)}$. In Sec.~\ref{Sec.:Add_result_figs} we also provide plots of droplets corresponding to individual ranks. The separable state used here as an example for tomography has also been used as an example for visualization in Fig.~\ref{fig:intro_drops}. 
\begin{table}[h]
	\caption{
		Experimental state fidelites ($\mathcal{F}_s$) corresponding to desired two-qubit quantum state $|\psi_t\rangle$, see Fig.~\ref{fig:BellNew} and \ref{fig:00+01New}.}
	\centering
	\begin{tabular}{p{4cm} p{3cm}}
		\hline
		\hline
		$|\psi_t\rangle$&$\mathcal{F}_s$\\[0.6ex]
		\hline
		$\frac{1}{\sqrt{2}}(|00\rangle+|11\rangle)$   &0.9989$\pm 1\times{10}^{-3}$\\[0.9ex]
		$\frac{1}{\sqrt{2}}(|00\rangle+|01\rangle)$   &0.9982$\pm 1\times{10}^{-3}$\\
		\hline
		\hline
	\end{tabular}
	\label{table: two qubit fid}
\end{table}

In the next section, we take this scanning tomography approach a step further and explain how this can be used to experimentally tomograph shapes representing unitary quantum processes. 
\section{Theory of Wigner quantum process tomography}
\label{WQPT}
In general, process tomography~\cite{nielsen2002quantum,QPT_nielsen,Childs_QPT} is a method to characterize a quantum process based on experimentally measured data. Here we focus on the tomography of unitary processes and translate the previously introduced Wigner quantum process tomography of known propagators~\cite{leiner2018wigner} in ensemble quantum devices to the setting of pure state near-term quantum devices. The considered unitary processes could refer to quantum gates, time evolution operators, or pulse sequences. \\

As shown in Sec.~\ref{Wigner_QST}, if the operator of interest is a quantum state or a density operator represented by $\rho^{[N]}$, the spherical droplet function $f_{j}^{(\ell)}(\beta,\alpha)$ can be measured experimentally. In the case of process tomography, our operator of interest is an $N$ qubit quantum process $U^{[N]}$. It is possible to scan the Wigner representation of an arbitrary operator $A^{[N]}$ if it can be experimentally mapped onto the density operator. In the next section, we present the algorithm for Wigner process tomography based on a method to map a unitary process matrix onto a density matrix~\cite{leiner2018wigner,Fahmy1,Fahmy2,Fahmy3}.
\subsection{Mapping of a unitary process matrix onto a density matrix}
\label{mapping}
Mapping a general \textit{unitary} matrix $U^{[N]}$ onto a \textit{Hermitian} density matrix of the same dimension is not possible. Here, we double the dimension of the density matrix by using an additional ancilla qubit $q_0$ and by a controlled process $cU^{[N+1]}$ operation we inscribe the unitary $U^{[N]}$ (and its adjoint $(U^{[N]})^{\dagger}$) in an off-diagonal block of the density matrix $\rho^{[N+1]}$ as shown below. Under $cU^{[N+1]}$, the unitary $U^{[N]}$ acts only on the target qubits $\textit{q}_{1},\dots,\textit{q}_{N}$ if the control qubit $\textit{q}_{0}$ is in state $\ket{1}$. The corresponding matrix representation of the controlled process $cU^{[N+1]}$ is
\begin{equation}
	\label{eq.31}
	{cU}^{[N+1]} = 
	\begin{pmatrix} 
		\mathds{1}^{[N]} 
		& 
		0^{[N]}  \cr
		0^{[N]}  & U^{[N]}
	\end{pmatrix},
\end{equation} 
where the top diagonal block corresponds to a $2^{N}\times2^{N}$ dimensional identity matrix $\mathds{1}^{[N]}$ and the lower diagonal block is the unitary $U^{[N]}$. The off-diagonal blocks are $2^{N}\times2^{N}$ dimensional zero matrices. \\

As shown in~\cite{leiner2018wigner} for ensemble quantum processors, $U^{[N]}$ can be mapped onto the density operator $\rho^{[N+1]}$ by preparing the ancilla (control) qubit $\textit{q}_{0}$ in the superposition state $\frac{1}{\sqrt{2}}(|0\rangle+|1\rangle)$ and the remaining system qubits $\textit{q}_{1},\dots, \textit{q}_{N}$ in the maximally mixed state. Hence, the prepared density operator is
\begin{equation}
	\label{eq.32}
	\rho_{0}^{[N+1]} = \frac{1}{2} \big(|{0}\rangle+|{1}\rangle\big)\big(\langle{0}|+\langle{1}|\big)  \otimes \frac{1}{2^{N}} (\mathds{1}^{[N]}),
\end{equation} 
and the density operator after application of $cU^{[N+1]}$ is
\begin{equation}
	\label{eq.33}
	\rho_{U}^{[N+1]} = {cU}^{[N+1]}\rho_{0}^{[N+1]}({cU}^{[N+1]})^\dagger, 
\end{equation} 
which can be rewritten in block matrix form as
\begin{equation}
	\label{eq.34}
	\rho_{U}^{[N+1]} = \frac{1}{2^{N+1}}
	\begin{pmatrix} 
		\mathds{1}^{[N]} 
		& 
		(U^{[N]})^{\dagger}  \cr
		U^{[N]}  & \mathds{1}^{[N]} 
	\end{pmatrix}.
\end{equation}
Using this approach, the unitary $U^{[N]}$ is now imprinted onto the density operator $\rho^{[N+1]}$ of the augmented system. Since the experimental implementation of a controlled process $cU^{[N+1]}$ requires the knowledge of $U^{[N]}$, this version of Wigner process tomography described here is in general only applicable for \textit{known} processes~\cite{leiner2018wigner}. 
\subsection{Wigner quantum process tomography}
\label{sec.WQPT}
\begin{figure}[h]
	\centering
	\includegraphics[scale=0.8]{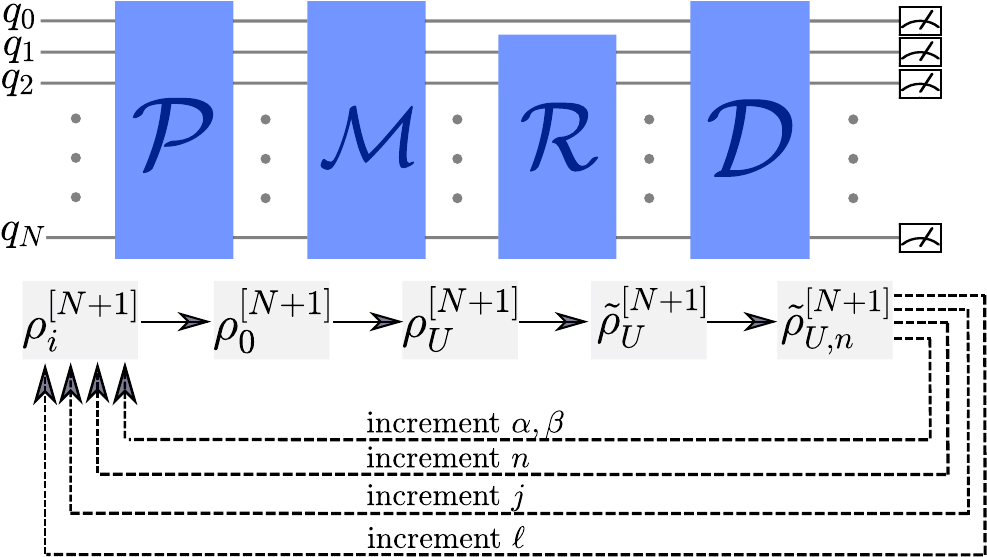}
	\caption{Schematic of the presented Wigner tomography algorithm for known unitary processes. The algorithm consists of four key blocks, namely Preparation ($\mathcal{P}$), Mapping ($\mathcal{M}$), Rotation ($\mathcal{R}$), and Detection-associated rotations ($\mathcal{D}$) followed by measurements. The rotation ($\mathcal{R}$) block acts only on system qubits $\textit{q}_{1},\dots, \textit{q}_{N}$, whereas all the other three blocks act on all the qubits $\textit{q}_{0},\textit{q}_{1},\dots, \textit{q}_{N}$. The lower part of the figure shows the evolution of the density matrix after each block. The algorithm is repeated for all desired combinations of parameters.}
	\label{fig:QPT_algo}
\end{figure}

Here, we first present the algorithm for process tomography and then explain each step individually for pure-state quantum processors. A droplet function $f_j^{(\ell)}$ representing a quantum process $U^{[N]}$ can be experimentally measured using the following steps (see Fig.~\ref{fig:QPT_algo}):
\begin{enumerate}[label=(\roman*)]
	\item \textbf{Preparation ($\mathcal{P}$)}: Prepare ancilla qubit $\textit{q}_{0}$ in the superposition state $\frac{1}{\sqrt{2}}(|0\rangle+|1\rangle)$ and effectively create the maximally mixed state of the system qubits $\textit{q}_{1},\dots, \textit{q}_{N}$ by temporal averaging.   
	\item \textbf{Mapping ($\mathcal{M}$)}: Implement the $cU^{[N+1]}$ operation to map the process $U^{[N]}$ onto $\rho_{U}^{[N+1]}$.  
	\item \textbf{Rotation ($\mathcal{R}$)}: Rotate the system qubits  $\textit{q}_{1},\dots, \textit{q}_{N}$ inversely for scanning.
	\item \textbf{Detection-associated rotations ($\mathcal{D}$)}: Apply local unitary operations to measure required expectation values of Pauli operator components of axial tensor operators $T_{j0}^{(\ell)[N]}$ (see Table~\ref{Tab:AxialTensors}) that are not directly measurable. 
\end{enumerate}
These four steps are repeated for a set of angles $\beta\in [0,\pi]$ and $\alpha\in [0,2\pi]$ and for different $n$, rank $j$ and labels $\ell$ to calculate the droplet function $f_{j}^{(\ell)}(\beta,\alpha)$. Now we elaborate each step individually. 
\\

\textit{Step 1}: The algorithm starts with the initial state $\rho_{i} = |0\dots0\rangle\langle0\dots0|$. The state $\rho_{0}$ (Eq.~\ref{eq.32}) is prepared by applying a Hadamard gate to qubit $q_0$ to achieve an equal superposition. The maximally mixed state of qubits $q_1,\dots,q_{N}$ is created by temporally averaging experiments for all the possible computational basis states by applying appropriate local NOT gates (see supplementary Sec.~\ref{sec.temporal_avg}). We discuss this in detail for a single-qubit system in Sec.~\ref{WQPT_expt}.\\

\textit{Step 2}: The operation $cU^{[N+1]}$ used for mapping can be experimentally implemented by decomposing it into elementary gates~\cite{Barenco_element} or using pulse-level control methods~\cite{KHANEJA2005296,Devra2018}. \\ 

\textit{Step 3}: Since our operator of interest is an $N$-qubit unitary process matrix $U^{[N]}$, Eq.~\ref{eq.9a} takes the form
\begin{equation}
	\label{eq.38}
	f_{j}^{(\ell)}(\beta,\alpha) = s_{j}\langle{T_{j,\alpha\beta}^{(\ell)[N]}}|{U^{[N]}}\rangle.
\end{equation}     
As shown in supplementary Sec.~\ref{WQPT_supp. sec.}, Eq.~\ref{eq.38} can be rewritten as
\begin{equation}
	\label{eq.39}
	f_{j}^{(\ell)}(\beta,\alpha) = s_{j}\langle{\sigma^{+}\otimes T_{j,\alpha\beta}^{(\ell)[N]}}\rangle_{\rho_{U}^{[N+1]}},
\end{equation} 
Similarly as in state tomography (Sec.~\ref{Wigner_QST}), instead of rotating the axial tensor operator $T_{j0}^{(\ell)[N]}$, we equivalently rotate the density matrix of the system qubits $\textit{q}_{1},\dots, \textit{q}_{N}$ inversely, such that:
\begin{equation}
	\label{eq.40}
	f_{j}^{(\ell)}(\beta,\alpha) = s_{j}\langle{\sigma^{+}\otimes T_{j0}^{(\ell)[N]}}\rangle_{\tilde{\rho}_{U}^{[N+1]}},
\end{equation} 
where
\begin{equation}
	\label{eq.41}
	\tilde{\rho}_{U}^{[N+1]} = (R_{\alpha\beta}^{[N+1]})^{-1}\rho_{U}^{[N+1]}R_{\alpha\beta}^{[N+1]},
\end{equation}
and $R_{\alpha\beta}^{[N+1]} = \mathds{1}^{[1]}\otimes R_{\alpha\beta}^{[N]}$ which corresponds to the rotation of only the system qubits $\textit{q}_{1},\dots, \textit{q}_{N}$ for scanning. Using the relation $\sigma^{+} = \frac{1}{2}(\sigma_{x}+i\sigma_{y})$, Eq.~\ref{eq.40} can be rewritten in terms of Pauli operators as:
\begin{equation}
	\label{eq.44}
	\begin{aligned}
		f_{j}^{(\ell)}(\beta,\alpha)  = {} & \dfrac{s_{j}}{2}\big(\langle{\sigma_{x} \otimes T_{j0}^{(\ell)[N]}}\rangle_{{\tilde{\rho}}_{U}^{[N+1]}} + \\ 
		& i\langle{\sigma_{y} \otimes T_{j0}^{(\ell)[N]}}\rangle_{{\tilde{\rho}}_{U}^{[N+1]}}\big).
	\end{aligned}
\end{equation}\\

\textit{Step 4}: In analogy to the case of Wigner state tomography, the expectation values of Pauli operators, which are not directly observable, can be measured with the help of local unitary operations $u_{n}$ (detection-associated rotations). 
\subsection{Estimation of unitary process matrices from droplet functions}
\label{sec.process_estimation}
Similar to the estimation of density matrices in the case of Wigner state tomography, unitary process matrices can also be estimated from the experimentally tomographed droplets. A general $N$-qubit unitary process matrix can be expressed in terms of Pauli operators as:
\begin{equation}
	\label{eq.45}
	U^{[N]} = \sum_{a=0}^3\sum_{b=0}^3\dots\sum_{g=0}^3 c_{ab\dots g} 	(\sigma_{a}\otimes\sigma_{b}\otimes\dots\otimes\sigma_{g}),
\end{equation}  
where $\sigma_{0}$ is a 2$\times$2 identity ($\mathds{1}$) matrix, while $\sigma_{1}$, $\sigma_{2}$ and $\sigma_{3}$ are the standard Pauli matrices $\sigma_{x}$, $\sigma_{y}$ and $\sigma_{z}$, respectively. The complex coefficients $c_{ab\dots g}$ can be computed by calculating the scalar product between basis droplets (ideally simulated without noise) and experimental droplets, as shown in Eq.~\ref{eq.18}. The basis droplets can be generated using the definitions provided in supplementary Sec.~\ref{Sec.:Sph_basis}. Based on the estimated process matrix $U^{[N]}$, the process tomography fidelity $\mathcal{F}_U$~\cite{glaser_unitaryControl} can be calculated using the relation: 
\begin{equation}
	\label{eq.46}
	\mathcal{F}_U = \dfrac{|\text{tr}(U^{[N]}(U^{[N]}_{t})^{\dagger})|}{2^{N}},
\end{equation}
where $U^{[N]}_{t}$ is a target unitary process matrix. 
\section{Experimental implementation of Wigner process tomography}
\label{WQPT_expt}
This section describes the approach to implementing the above-mentioned Wigner process tomography on experimental quantum devices. Here, we present the simulated and experimental process tomography results performed on IBM quantum devices for a pure state of an individual quantum system. The quantum circuits provided here are general and can be directly adapted to other near-term quantum devices. The Wigner process tomography can be directly implemented using the Python-based software package \texttt{DROPStomo}~\cite{DROPStomo} for a single-qubit system.

\subsection{Single qubit system}
\label{WQPT_expt_single_qubit}
For the Wigner process tomography of a single-qubit ($N=1$) system, the total number of qubits required is two ($q_0$ and $q_1$), where $q_0$ is an ancilla qubit and $q_1$ is the system qubit. For a single-qubit system, the possible values of rank $j$ are (c.f. Table \ref{Tab:AxialTensors}): $j=0$ for label $\ell=\emptyset$, and $j=1$ for label $\ell=1$. Hence, a single-qubit unitary process is represented by the spherical functions $f_{0}^{(\emptyset)}$ and $f_{1}^{(1)}$, which can be calculated based on the measured expectation values of Eq.~\ref{eq.44} as:
\begin{equation}
	\label{eq.47}
	\begin{aligned}
		f_{0}^{(\emptyset)}(\beta,\alpha) = {} & \dfrac{1}{2}\sqrt{\dfrac{1}{4\pi}}\big(\langle{\sigma_{x} \otimes T_{00}^{(\emptyset)[1]}}\rangle_{{\tilde{\rho}}_{U}^{[2]}}\\
		& + i \langle{\sigma_{y} \otimes T_{00}^{(\emptyset)[1]}}\rangle_{{\tilde{\rho}}_{U}^{[2]}}\big)\\
		f_{1}^{(1)}(\beta,\alpha) = {} & \dfrac{1}{2}\sqrt{\dfrac{3}{4\pi}}\big(\langle{\sigma_{x} \otimes T_{10}^{(1)[1]}}\rangle_{\tilde{{\rho}}_{U}^{[2]}}\\
		& + i \langle{\sigma_{y} \otimes T_{10}^{(1)[1]}}\rangle_{\tilde{{\rho}}_{U}^{[2]}}\big).  	
	\end{aligned}
\end{equation} 
Substituting the explicit form of the tensor operators $T_{00}^{(\emptyset)}$ and $T_{10}^{(1)}$ from Table \ref{Tab:AxialTensors} gives
\begin{equation}
	\label{eq.48}
	\begin{aligned}
		f_{0}^{(\emptyset)}(\beta,\alpha) & = \dfrac{1}{4}\sqrt{\dfrac{1}{2\pi}}\big(\langle{\sigma_{0x}}\rangle_{\tilde{{\rho}}_{U}^{[2]}} + i \langle{\sigma_{0y}}\rangle_{{\tilde{\rho}}_{U}^{[2]}}\big)\\
		f_{1}^{(1)}(\beta,\alpha) & =  
		\dfrac{1}{4}\sqrt{\dfrac{3}{2\pi}}\big(\langle{\sigma_{0x}\sigma_{1z}}\rangle_{{\tilde{\rho}}_{U}^{[2]}} + i \langle{\sigma_{0y}\sigma_{1z}}\rangle_{{\tilde{\rho}}_{U}^{[2]}}\big).
	\end{aligned}
\end{equation} 
We first focus on the preparation step ($\mathcal{{P}}$) of the algorithm, i.e., preparing qubits $\textit{q}_{0}$ and $\textit{q}_{1}$ in a state whose density matrix corresponds to Eq.~\ref{eq.32} for $N=1$. The preparation of the ancilla qubit $q_0$ in the superposition state can be achieved straightforwardly by applying the Hadamard (H) gate on $q_0$. A relatively straightforward approach to preparing the system qubit in the maximally mixed state would be to prepare it in the state $|0\rangle$ and to repeat each experiment multiple times, where, in each repetition, it is randomly decided (with 50\% probability) whether to flip the qubit using a NOT gate. However, the standard deviation of the average population of state $|1\rangle$ from the expected value of 0.5 decreases only with the inverse of the square root of the number of repetitions in this probabilistic approach. Hence, a large number of repetitions would be required to minimize this additional noise source due to the imperfect realization of the completely mixed state. For example, to achieve a standard deviation of less than 0.005, about 8000 repetitions would be necessary, even in the case of perfect gate operation. In contrast, the completely mixed state can be created exactly using a temporal averaging approach~\cite{knill_temporal,preskill1998lecture}, where the experiment is only repeated for the set of computational basis states. In the case of a single system qubit, this requires only two experiments: one experiment without a flip of initial state $|0\rangle$ and one experiment with a flip, see supplementary Sec.~\ref{sec.temporal_avg}.\\

Interestingly, temporal averaging was initially introduced in quantum information processing to mimic experiments of a pure state by averaging expectation values obtained by measuring a set of experiments on (partially) mixed states of an ensemble quantum processor~\cite{knill_temporal}. In contrast, here experiments on the maximally mixed state of the system qubits $\textit{q}_{1},\dots, \textit{q}_{N}$ are mimicked by averaging expectation values obtained by measuring a set of experiments with pure states.        
\begin{figure}[h]
	\includegraphics[scale=0.85]{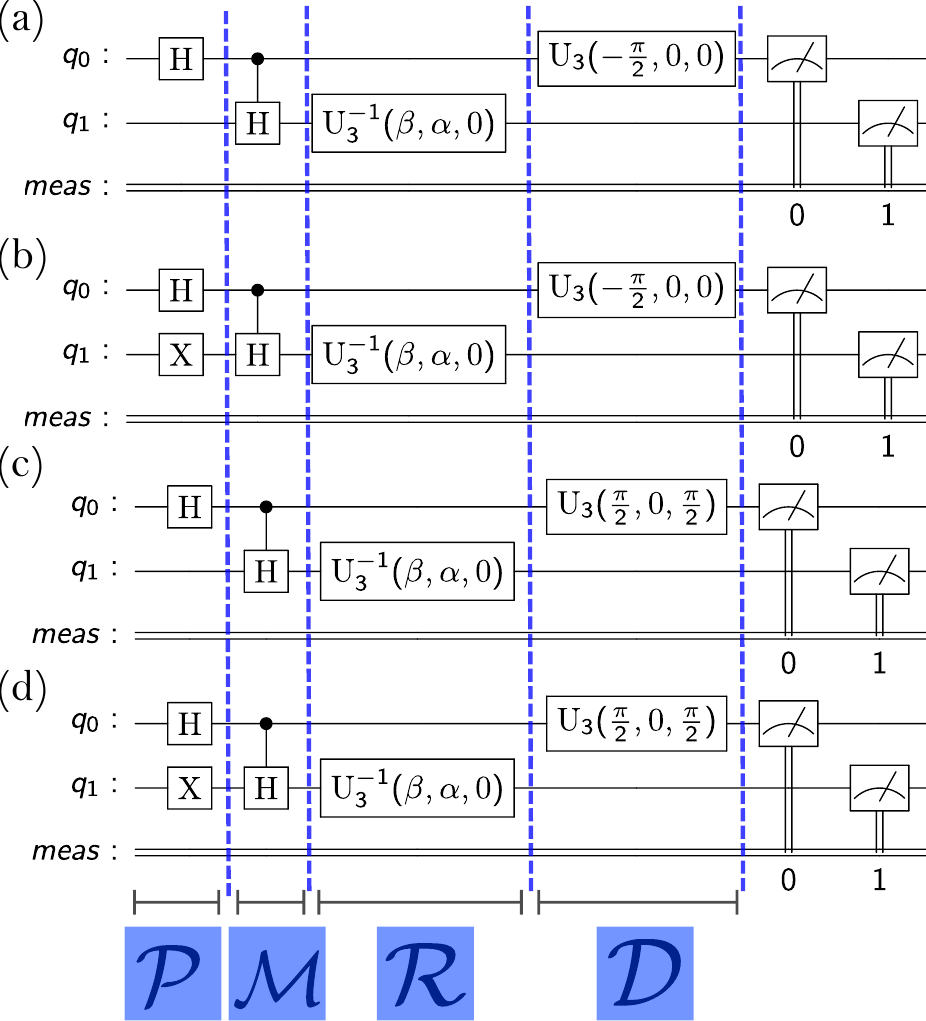}
	\caption{Set of quantum circuits for the Wigner process tomography of the Hadamard (H) gate. The initial state of the qubits is $|\psi\rangle_i = |00\rangle$. The four blocks of the algorithm preparation ($\mathcal{P}$), mapping ($\mathcal{M}$), rotation ($\mathcal{R}$), and detection-associated rotations ($\mathcal{U}$) are explicitly shown here. The $\mathrm{U}_{3}$ gate used in the circuit is discussed in supplementary Sec.~\ref{supp:U3}.}
	\label{fig:QPTCircuit}
\end{figure}\\

As an example, in Fig.~\ref{fig:QPTCircuit} we present the set of quantum circuits for process tomography where the unitary operator of interest is the Hadamard gate (H). Hence, a controlled Hadamard gate is applied in the mapping step ($\mathcal{M}$) of each experiment in (a)-(d). After the preparation step $\mathcal{P}$, the system qubit $q_1$ is in state $|0\rangle$ for circuits (a) and (c), whereas it is switched to $|1\rangle$ by applying a NOT (X) gate on $q_1$ in circuits (b) and (d). The (temporally) averaged expectation values of experiments (a) and (b) provide $\langle{\sigma_{0x}}\rangle$ (and also $\langle{\sigma_{0x}\sigma_{1z}}\rangle$). Similarly, the (temporally) averaged expectation values of experiments (c) and (d) provide $\langle{\sigma_{0y}}\rangle$ (and also $\langle{\sigma_{0y}\sigma_{1z}}\rangle$). Using Eq.~\ref{eq.48}, these expectation values can be combined to obtain the droplet functions $f_0^{(\emptyset)}$ and $f_1^{(1)}$, see Fig~\ref{fig:QPT_results} (first column) for the combined ($f = f_0^{(\emptyset)}+f_1^{(1)}$) droplets~\cite{leiner2018wigner} and Fig.~\ref{fig:QPT_results_ranks} (first row) for the individual droplets $f_0^{(\emptyset)}$ and $f_1^{(1)}$. Fig.~\ref{fig:QPT_results} also shows experimental droplets of the NOT (X) gate and the process corresponding to a rotation of $\frac{3\pi}{2}$ around the $y$ axis. The separate droplets $f_0^{(\emptyset)}$ and $f_1^{(1)}$ for rank $j=0$ and $j=1$ are provided in the supplementary Fig.~\ref{fig:QPT_results_ranks}. \\

Based on the experimentally measured droplet functions $f_0^{(\emptyset)}$ and $f_1^{(1)}$, a process matrix can be estimated (as shown in Sec.~\ref{sec.process_estimation}): any single-qubit unitary process can be expressed in terms of Pauli operators \cite{nielsen2002quantum} as
\begin{equation}
	\label{eq.49}
	U^{[1]} = \sum_{k=0}^{3}c_{k}\sigma_{k}
\end{equation}
with in general complex coefficients $c_k$ for $k\in\{0,1,2,3\}$. Using Eq.~\ref{eq.18}, the coefficients   $c_k$ are obtained by calculating the scalar product between the basis droplets $f_{\sigma_{k}}$ (refer to Sec.~\ref{Sec.:Basis_single}) and the sum of the experimentally tomographed droplets $f_{0}^{(\emptyset)}$ and $f_{1}^{(1)}$: 
\begin{equation}
	\label{eq.50}
	c_{k} = \langle{f_{\sigma_{k}}|f_{0}^{(\emptyset)}+f_{1}^{(1)}}\rangle.
\end{equation}
Table~\ref{table: QPT fid} summarizes the experimental process fidelities of the gates considered above. The standard deviation given in Table~\ref{table: QPT fid} were estimated by conducting the experiment three times for the NOT gate for a reference.
\begin{table}[h]
	\caption{
		Experimental process tomography fidelity ($\mathcal{F}_U$) corresponding to target quantum gates $U_t$. The corresponding droplets are shown in Fig.~\ref{fig:QPT_results}.}
	\centering
	\begin{tabular}{p{4.5cm} p{2.5cm}}
		\hline
		\hline
		$U_t$&$\mathcal{F}_U$\\[0.6ex]
		\hline
		Hadamard (H)   &$0.9506\pm1\times10^{-3}$\\[0.8ex]
		NOT (X)   & $0.9679\pm1\times10^{-3}$\\[0.8ex]
		$\big[\frac{3\pi}{2}\big]_{y}$ & $0.9407\pm1\times10^{-3}$\\
		\hline
		\hline
	\end{tabular}
	\label{table: QPT fid}
\end{table} 
\section{Understanding errors using Wigner state and process tomography}
\label{Sec.:DROPS_errors}
Quantum devices are prone to different kinds of errors both in the implementation of desired states and of quantum gates. Here, we focus on the example of rotation errors, e.g., due to errors in pulse calibrations, etc. Visualizing or identifying these errors directly is useful in quantum information processing. Here, we show how the DROPS representation is helpful to achieve this. As described in the caption of Fig.~\ref{fig:intro_drops}, the radius (distance from the origin to a point on the sphere) of a droplet represents the absolute value of a droplet function $f^{(\ell)}$, and color represents the phase $\varphi=\text{arg}[f^{(\ell)}]$. In addition, the direction of a qubit droplet reflects the direction of the Bloch vector for quantum states (see: Fig.~\ref{fig:intro_drops} and \ref{fig:QST result single qubit}) and of the rotation axis for single-qubit quantum gates (see Fig.~\ref{fig:QPT_results}).\\

\begin{figure}[h]
	\centering
	\includegraphics[scale=2]{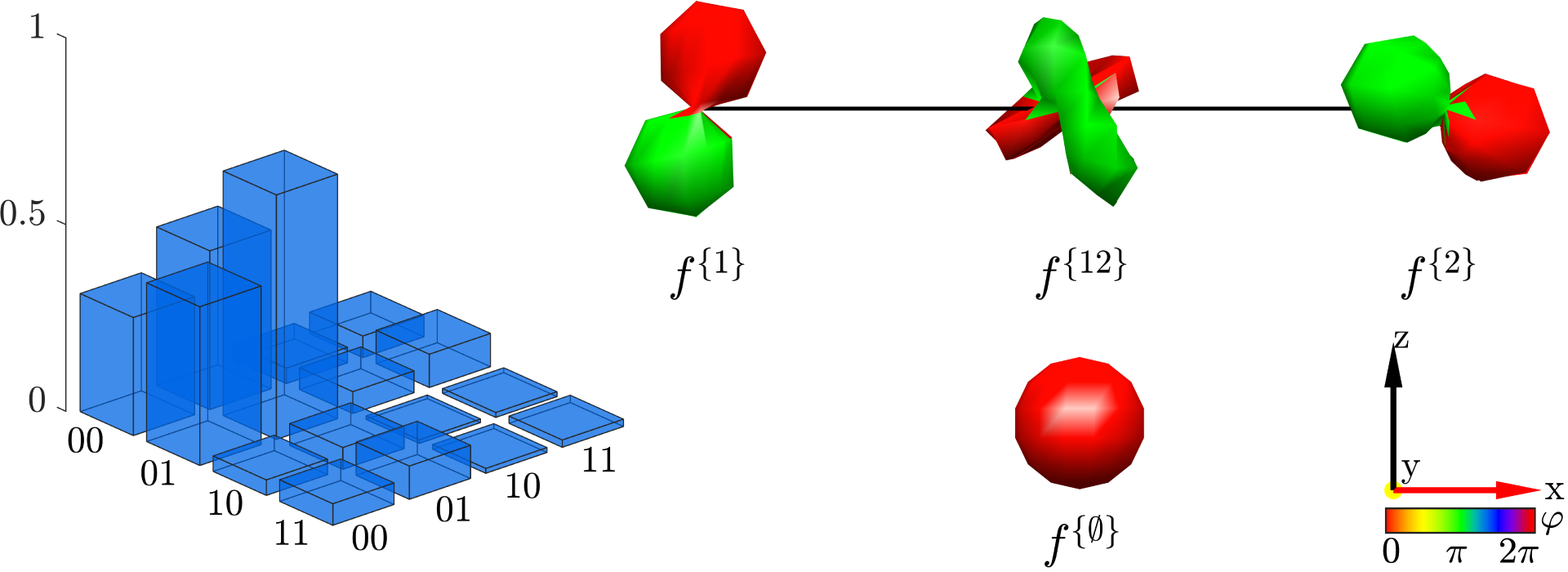}
	\caption{Skyscraper visualization of the real part of the density matrix (left) and the experimentally tomographed DROPS visualization of the full density matrix (right) corresponding to the desired state $|\psi\rangle = \frac{1}{\sqrt{2}}(|00\rangle+|01\rangle)$ with additional rotations of $\mathrm{U}_3 (\pi/12,0,0)$ on qubit $q_1$, and $\mathrm{U}_3 (\pi/9,\pi/12,0)$ on qubit $q_2$ in the preparation step. Refer to Sec.~\ref{supp:U3} for $\mathrm{U}_{3}$ gates.}
	\label{fig:Error}
\end{figure}

As an example, we show the experimental tomography result of the desired quantum state $|\psi\rangle = \frac{1}{\sqrt{2}}(|00\rangle+|01\rangle)$ with some error in the state preparation. We deliberately introduce an additional rotation of $\mathrm{U}_3 (\pi/12,0,0)$ on qubit $q_1$, and $\mathrm{U}_3 (\pi/9,\pi/12,0)$ on qubit $q_2$ in the preparation step. Fig.~\ref{fig:Error} shows that these kinds of errors are directly visible in the DROPS representations (on the right) as misalignment of the linear droplet functions $f^{(1)}$ and $f^{(2)}$ compared to the experimental tomography results of the case without rotation errors as shown in Fig.~\ref{fig:00+01New} and to the ideal case shown in Fig~\ref{fig:intro_drops}. Note that such a direct physical interpretation of the error terms is not possible using the standard skyscraper visualization~\cite{nielsen2002quantum} of the density matrix (on the left). In Fig.~\ref{fig:Error}, only the skyscraper visualization of the real part of the density matrix is shown. The imaginary part is plotted in Fig.~\ref{fig:Error_and_ideal} along with decomposed bilinear droplet functions for the state with and without rotation errors. In Fig.~\ref{fig:DiffView}, we also show the droplet plots from a different perspective to emphasize the misalignment errors.  
\section{Scanning using different sampling schemes: a numerical study}
\label{Sec.:sampling_techqniues}
As we have seen, scanning is a key step in the Wigner tomography approach, and hence choosing a suitable sampling scheme on a sphere is important. This is a topic of interest in the general field of signal processing~\cite{kennedy_sadeghi_2013}, and a number of different sampling schemes have been proposed in the literature. \\

In the ideal case of negligible experimental noise, only a small number of sampling points would be necessary to determine the correct expansion coefficients of spherical harmonics as each droplet function is band-limited~\cite{Khalid, leiner2018wigner}. An advantage of using a large number of sampling points is to obtain a direct view of the experimentally measured droplet shapes without additional signal processing steps, such as the extraction of expansion coefficients of spherical harmonics or the estimation of the matrix representation of an operator. Note that a larger number of sampling points $N_{p}$ does not necessarily entail an increase in the total number of experiments $N_{tot} (= N_{p}\cdot N_{s})$ because the number $N_{s}$ of shots per sampling point can also be adapted to each sampling scheme. In the following, we will compare the performance of different sampling schemes for (approximately) the same total number of shots ($N_{tot}$) given by the product of the number of sampling points ($N_{p}$) and the number of shots per sampling point  ($N_{s}$). \\

\begin{figure*}[ht]
	\centering
	\includegraphics[scale=1]{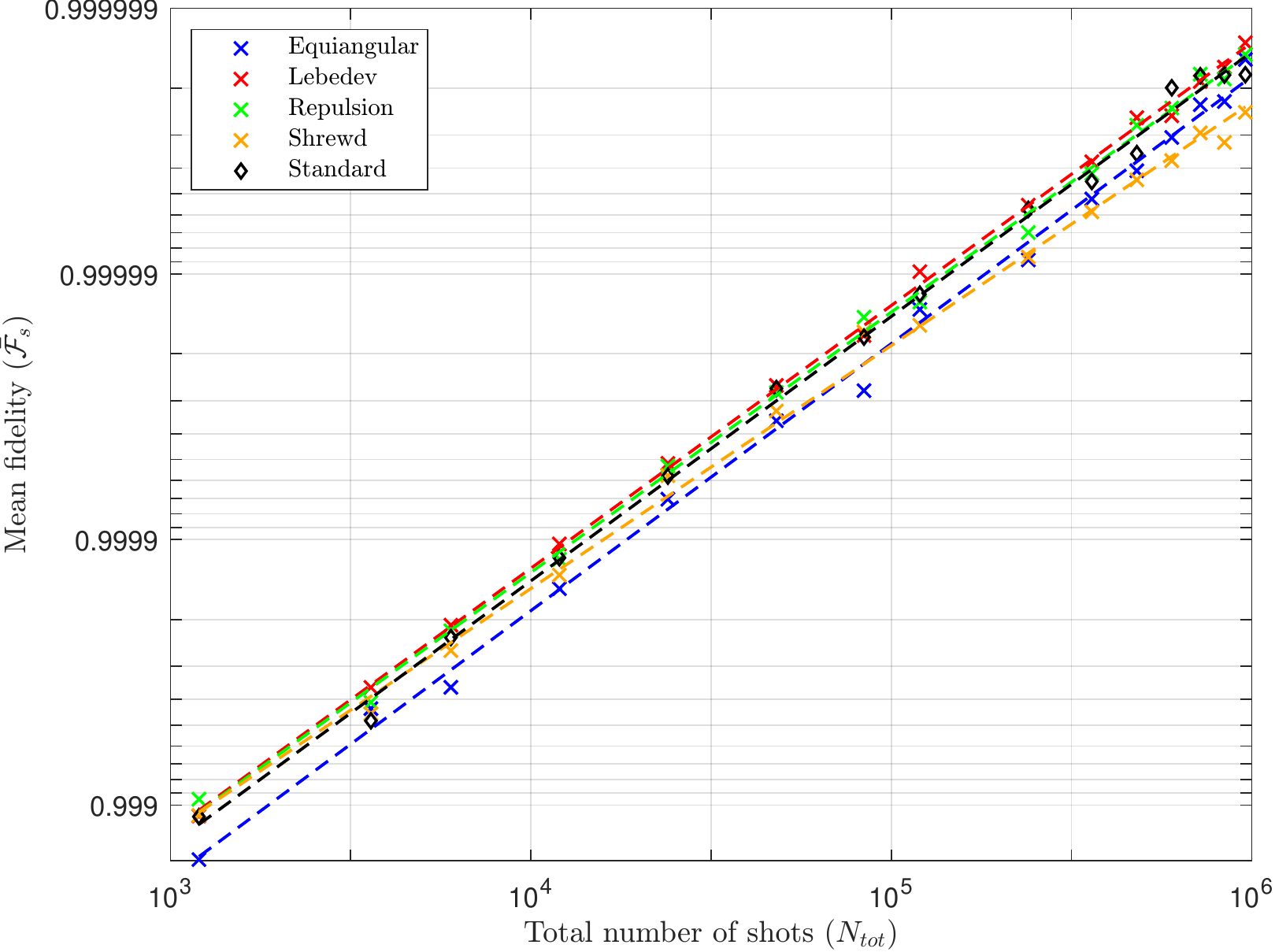}
	\caption{Plot of the mean fidelity ($\bar{\mathcal{F}_s}$) as a function of the total number of shots $(N_{tot})$ for different sampling techniques and for the standard state tomography method applied to the quantum state $|\psi\rangle = (-0.69-0.098i)|0\rangle+(0.66+0.30i)|1\rangle$. The mean fidelity is calculated by repeating the simulation 100 times for each data point. In the simulation, only the noise due to a limited number of shots is considered. The corresponding plot with standard deviation is available in supplementary Sec.~\ref{Sec.:shots_study}.}
	\label{fig:ShotSim}
\end{figure*}

In Fig.~\ref{fig:ShotSim} the mean fidelity ($\bar{\mathcal{F}_s}$) of the tomographed state is shown as a function of the total number of experiments for the Lebedev~\cite{LEBEDEV197610}, REPULSION~\cite{BAK1997132}, and SHREWD~\cite{EDEN1998220} sampling schemes along with the simple equiangular grid and the standard tomography method~\cite{nielsen2002quantum, PhysRevA.64.052312}. For more detailed information, including standard deviations, see supplementary Sec.~\ref{Sec.:shots_study}. We only consider the noise due to a limited number of shots. In the simple case of an equiangular grid~\cite{DRISCOLL1994202, 6006544} of eight polar angles $\beta\in\{0,\frac{\pi}{7},\dots\pi\}$ and fifteen azimuthal angles $\alpha\in\{0,\frac{2\pi}{14},\dots2\pi\}$ as shown in Fig.~\ref{fig:grid_plot}a, the total number of grid points is 120. For Lebedev, REPULSION, and SHREWD 110 grid points were used. Since the number of sampling points in both cases is similar, for simplicity, the same number of shots per sampling point was chosen. In contrast, for the standard tomography method, only three measurement settings are required for the case of a single-qubit. The forty-fold decrease in the number of sampling points was compensated by correspondingly increasing the number of shots per measurement setting by a factor of 40 to arrive at the same total number of shots $N_{tot}$ as in the previous cases. \\

In the supplementary Sec.~\ref{Sec.:shots_study}, we also provide results for the state
$|\psi\rangle =\frac{1}{\sqrt{2}}(|0\rangle+|1\rangle)$. The plots indicate that the mean fidelity differs depending on sampling techniques and tomography methods and also show that an equiangular grid is not an optimal choice, as expected because the density of grid points is higher near the poles compared to the equator of the sphere. This numerical study is expected to help make an informed choice of the sampling scheme for quantum computing applications in which an estimate of the quantum state with high precision is required.\\

We used the Spinach~\cite{spinach} software to generate angles and weights for the REPULSION and the SHREWD sampling techniques. For standard state tomography, the maximum-likelihood estimation (MLE) method \cite{smolin2012efficient,SINGH_valid} was used on the numerical data to estimate a valid quantum state. We used the corresponding qiskit~\cite{gadi_aleksandrowicz_2019_2562111} classes to perform the standard state tomography based on the MLE method. 

\section{\texttt{DROPStomo}: A Python-based software package for Wigner state and process tomography}
\label{Sec.:Code}
\texttt{DROPStomo}~\cite{DROPStomo} is a Python-based software package for performing Wigner state tomography for one- and two-qubit systems and process tomography for a single-qubit system. With \texttt{DROPStomo}, users can simulate (on a simulator or on quantum hardware) and analyze the tomographed droplets interactively. The package is based on the Qiskit framework~\cite{gadi_aleksandrowicz_2019_2562111}. However, it is straightforward to adapt it to other frameworks. \texttt{DROPStomo} can be installed and imported using the following command:
\begin{lstlisting}[language=python]
# install the package
pip install DROPStomo
# import the required modules
from DROPStomo import WQST1Q
from DROPStomo import WQST2Q
from DROPStomo import WQPT1Q
\end{lstlisting}
Here, we give a code snippet for performing Wigner state tomography for the one-qubit state $|\psi\rangle=\frac{|0\rangle+|1\rangle}{\sqrt{2}}$ for eight polar angles $\beta\in[0,\pi]$ and fifteen azimuthal angles $\alpha\in[0,2\pi]$. 

\begin{lstlisting}[language=python]
# state preparation gate
Up = U3Gate(theta=pi/2, phi=0, lam=0) 

# sampling for scanning
res_beta= 8
res_alpha = 15

# prepare quantum circuits
circ_q = WQST1Q.WQST_1Q_circuits(res_beta,Up)

# provide a simulator or a quantum hardware. For example:
simulator = Aer.get_backend('qasm_simulator')

# target density matrix
rho = np.matrix([[0.5,0.5],[0.5,0.5]])

# running quantum circuits on a simulator or quantum hardware.
WQST1Q.WQST_1Q_runner(res_beta,circuits=circ_q,qdevice=simulator,shots=8192,inter=1,rhoT=rho)

# ***Output***
# Experimental tomographed droplets (non-interactive if inter=0 and interactive if inter=1).
# Experimental expectation values.
# Experimental density matrix with state fidelity.






\end{lstlisting}
We provide the extended tutorial codes for two-qubit Wigner state tomography and one-qubit Wigner process tomography in our repository~\cite{DROPStomo}. 

\section{Discussion}
\label{Discussion}
In this work, we developed a general approach for Wigner tomography of quantum states and quantum processes for pure-state quantum devices with projective measurements by adapting the methodology described in~\cite{leiner2017wigner} and~\cite{leiner2018wigner}. We demonstrated the experimental implementation of these tomography approaches on IBM quantum devices. The experimentally measured droplet shapes provide a unique, vivid visual representation of abstract quantum operators, which reflects their essential features. For example, for a single-qubit, the droplet orientation provides essentially the same information as the Bloch vector representation. However, the DROPS representation is not limited to single-qubit systems but can also be used to visualize states and processes in multi-qubit systems, where different droplets provide information about the expectation values of multi-qubit operators, such as $\sigma_{1x}\sigma_{2x}$ or $\sigma_{1y}\sigma_{2z}\sigma_{3y}$ etc. The presented approach has similar limitations with respect to the number of qubits due to the exponential growth of the Hilbert space dimension as conventional tomography methods. However, for a small number of qubits, Wigner-type DROPS tomography forms an easily implementable alternative approach with additional benefits and without any additional cost in terms of experimental time and signal-to-noise ratio. In particular, the DROPS visualization allows one to directly see the kind of errors present in a given realization of quantum states and processes.\\

Originally, Wigner state and process tomography was developed for ensemble quantum processors, such as NMR~\cite{leiner2017wigner,leiner2018wigner}. The main purpose of this paper was to show that it is indeed possible to also apply it to standard quantum devices based on pure states. To achieve this, the following points have been explicitly addressed and discussed: 
\begin{enumerate}[label=(\alph*)]
	\item The description of DROPS tomography was rewritten using the language of quantum information processing instead of the language of NMR. A simple example is the consistent use of Pauli operators $\sigma_{x}$, $\sigma_{y}$ and $\sigma_{z}$ instead of the corresponding spin operators $I_{x}$, $I_{y}$, and $I_{z}$, which are not only unfamiliar outside of the NMR community but also differ by a factor of two. Another example is the description of quantum operations in terms of elementary quantum gates using the QASM~\cite{qasm} nomenclature instead of their descriptions in terms of rotations and pulse sequences.  
	\item Whereas it is natural to measure expectation values directly on ensemble quantum processors, in pure-state quantum processors, expectation values are typically estimated by averaging the outcomes of projective measurements for many repetitions of the experiment. The measurement of expectation values of single-qubit and multi-qubit operators necessary for DROPS tomography is explicitly discussed.
	\item The fact that DROPS tomography of unitary processes requires an ancilla qubit to be prepared in the \textit{completely mixed state} could create the false impression that it cannot be applied to pure-state quantum processors. We removed this hurdle by explaining and demonstrating how the concept of temporal averaging can be used to circumvent this problem. 
	\item We also showed how to implement discretized scalar products between droplet functions defined on a finite number of sampling points and how to use them to extract the standard matrix representation and the fidelity of states and processes based on experimentally measured droplets. Furthermore, we presented the results of a numerical study of the effect of different sampling schemes on the fidelity with which states can be experimentally tomographed.
	\item Finally, for a convenient adaption of the presented approaches, we provided the Python package \texttt{DROPStomo}~\cite{DROPStomo} for a direct implementation using Qiskit, which can also be adapted to other frameworks in a straightforward way.
\end{enumerate}
  
\acknowledgments
This project has received funding from the European Union’s Horizon 2020 research and innovation program under the Marie-Sklodowska-Curie grant agreement No 765267 (QuSCo). S.G. acknowledges funding by the German Research Foundation (DFG) under Germany’s Excellence Strategy – EXC-2111 – 390814868.  D.H. acknowledges support from the Verband der chemischen Industrie e.V (VCI). The project is part of the Munich Quantum Valley (MQV) initiative, which is supported by the Bavarian state government with funds from the Hightech Agenda Bayern Plus. We thank Frederik vom Ende for his useful comments on the manuscript. We acknowledge the use of IBM Quantum services for this work. The views expressed are those of the authors, and do not reflect the official policy or position of IBM or the IBM Quantum team. The quantum circuits presented in this paper were prepared using the latex package Q-circuit~\cite{eastin2004q}.

%
\bibliographystyle{unsrtnat}
\bibliography{bibfile}


\onecolumn
\clearpage
\begin{center}
	\textbf{\large Supplemental Materials}
\end{center}
\setcounter{section}{0}
\renewcommand{\thesection}{S-\Roman{section}}
\renewcommand{\thesubsection}{S-\Roman{section}.\Roman{subsection}}
\setcounter{subsection}{0}
\setcounter{equation}{0}
\setcounter{figure}{0}
\setcounter{table}{0}
\setcounter{page}{1}
\makeatletter
\renewcommand{\theequation}{S\arabic{equation}}
\renewcommand{\thefigure}{S\arabic{figure}}
\renewcommand{\bibnumfmt}[1]{[S#1]}
\section{Visualization of operators}
\label{Supp:visualization}
A method for representing and visualizing arbitrary quantum operators such as density matrices, Hamiltonians, unitary processes, etc., was proposed in~\cite{DROPS_main}, which is also known as DROPS representation (Discrete Representation of OPeratorS). Although the DROPS representation can be applied to arbitrary finite-dimensional quantum systems~\cite{DROPS_main,leiner2020symmetry}, here we focus on systems consisting of one or more qubits which are of particular interest in quantum information processing. \\

\begin{figure}[h]
	\centering
	\includegraphics[scale=0.9]{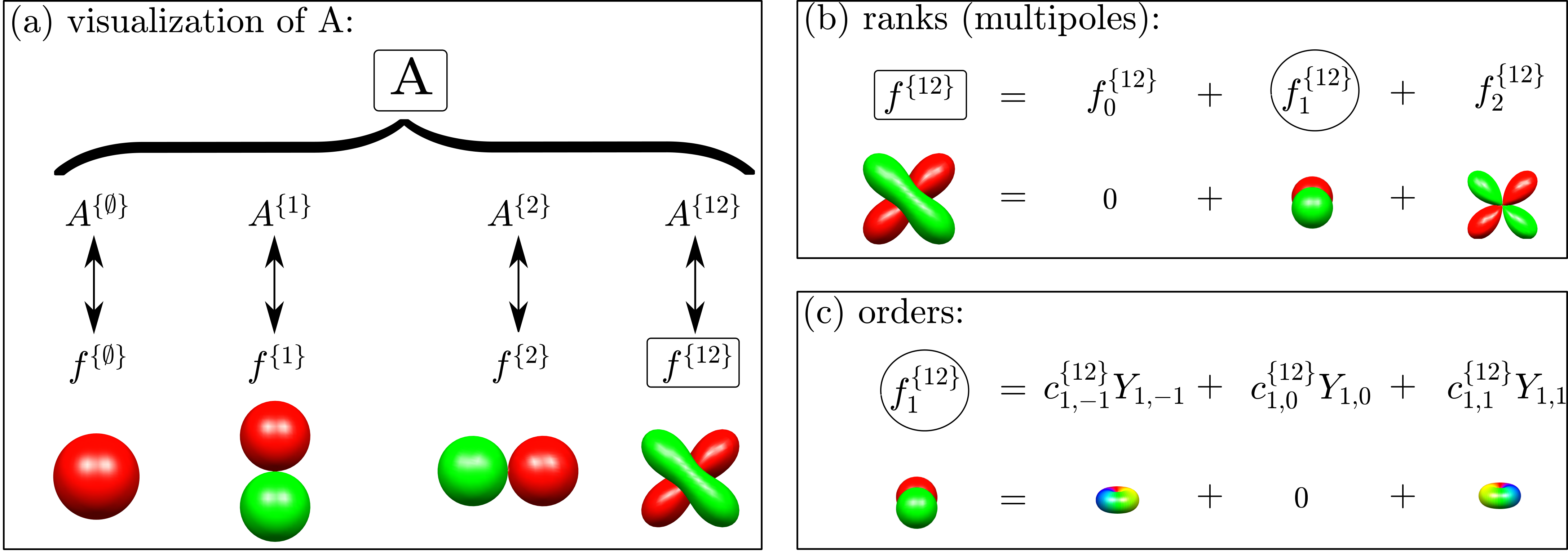}
	\caption{(a) For the two-qubit state $\frac{1}{\sqrt{2}}(|00\rangle+|01\rangle)$, the density operator is given by $A=\frac{1}{4}(\mathds{1}+\sigma_{1z}+\sigma_{2x}+\sigma_{1z}\sigma_{2x})$ and is visualized using multiple spherical functions $f^{\{\ell\}} = f^{\{\ell\}}(\beta,\alpha)$. The individual droplet operators $A^{\{\ell\}}$ of operator $A$ are mapped to the spherical droplet functions $f^{\{\ell\}}$. (b) $f^{\{12\}}$ (in box) is decomposed into its contributions $f^{\{12\}}_{j}$ with $j\in\{0,1,2\}$. (c) $f^{\{12\}}_{1}$ (in circle) is decomposed into spherical harmonics of order $m\in\{-1,0,1\}$.} 
	\label{suppfig:DROPS_example}
\end{figure}

The method maps an arbitrary operator $A$ on a set of spherical functions $f^{(\ell)}$ which can easily be visualized, e.g.  in the polar representation used here (or alternatively as colored spherical surfaces)~\cite{leiner2017wigner}. Panel (a) of Fig.~\ref{suppfig:DROPS_example} shows an example where the density operator $A$ of a two-qubit state is mapped to a set of four individual droplet functions $f^{(\ell)}$.
For systems consisting of up to two qubits, each of the individual droplet functions  ($f^{\{\emptyset\}}$, $f^{\{  1 \}}$,  $f^{\{  2 \}}$, and $f^{\{  12 \}}$)
simply corresponds to one of the four possible (sub) sets of qubits: $\{\emptyset \}$ labels the empty set,   $\{  1 \}$ labels the subset consisting only of the first qubit,  $\{  2 \}$ labels the subset consisting only of the second qubit, and $\{  12 \}$ labels the set consisting of both qubits.
Each subset is specified by a corresponding superscript label $\ell$. (For systems consisting of more than two qubits, specifying the (sub) systems of qubits is not sufficient and additional selection criteria, such as permutation symmetry, are necessary to specify each droplet~\cite{DROPS_main}.)
The method first decomposes any given operator $A$ as a sum of operators $A^{(\ell)}$, which are defined based on the criteria of the labels $\ell$ specified above. For example, the operator $A^{\{  1 \}}$ acts only on the first qubit, whereas, e.g., the operator $A^{\{  12 \}}$ acts on both the first and the second qubit etc.\\

As shown on  the left side of the double-headed arrow in Eq.~\ref{eq.1} (\textit{vide infra}), each droplet operator $A^{(\ell)}$ can be expanded in an operator basis consisting of irreducible spherical tensor operator components $T^{(\ell)}_{jm}$ with expansion coefficients $c_{jm}^{(\ell)}$~\cite{DROPS_main,leiner2020symmetry}, where  $j$ is the rank and $m$ is the order of the  spherical tensor operators $T^{(\ell)}_{jm}$. Based on the well-known correspondence~\cite{silver2013irreducible} between irreducible tensor operator components $T_{jm}$ and spherical harmonics $Y_{jm}$, a bijective mapping between the droplet operators $A^{(\ell)}$ and corresponding spherical droplet functions $f^{(\ell)}$ can be defined as
\begin{equation}
	\label{eq.1}
	A^{(\ell)} = \sum_{j \in J(\ell)}             \sum_{m=-j}^{j} c_{jm}^{(\ell)} T^{(\ell)}_{jm}                 \longleftrightarrow f^{(\ell)} = \sum_{j \in J(\ell)}                     \sum_{m=-j}^{j} c_{jm}^{(\ell)} Y_{jm},
\end{equation}
where identical expansion coefficients $c_{jm}^{(\ell)}$  are used on both sides  of the double-headed arrow.
Eq.~\ref{eq.1} can be rewritten in the more compact form
\begin{equation}
	\label{eq.2}
	A^{(\ell)} = \sum_{j \in J(\ell)}             A^{(\ell)}_{j}                  \longleftrightarrow f^{(\ell)} = \sum_{j \in J(\ell)}                      f^{(\ell)}_{j},
\end{equation}
where we defined the rank-$j$ droplet operators $A^{(\ell)}_{j}$ and droplet functions $f^{(\ell)}_{j}$ as
\begin{equation}
	\label{eq.3}
   A^{(\ell)}_{j} = \sum_{m=-j}^{j} c_{jm}^{(\ell)} T^{(\ell)}_{jm}   \ \ \ \ {\rm and} \ \ \ \       f^{(\ell)}_{j} = \sum_{m=-j}^{j} c_{jm}^{(\ell)} Y_{jm}.
\end{equation}
As shown for the example of the droplet  $f^{\{  12 \}}$ in panel (b) of Fig.~\ref{suppfig:DROPS_example}, each spherical droplet function  $f^{(\ell)}$  can be expressed as a sum of spherical functions $f^{(\ell)}_j$ with different ranks $j$.
For the case of $f^{\{  12 \}}$, the rank $j$ can be  0, 1, or 2. (Note that in the special case of the operator A represented in Fig. S1, the droplet function with rank 0 happens to vanish, i.e., $f^{\{  12 \}}_0=0$.) This graphically illustrates the decomposition given on the right side of the double-headed arrow in Eq.~\ref{eq.2}.

The decomposition of $ f^{(\ell)}_{j}$ in terms of the spherical harmonics $Y_{jm}$ with the expansion coefficients $c_{jm}^{(\ell)}$ (see right hand side of Eq.~\ref{eq.3}) is illustrated
in panel (c) of Fig.~\ref{suppfig:DROPS_example} for the rank $j=1$ spherical function  $f^{\{  12 \}}_1$, which is decomposed in terms of the spherical harmonics $Y_{1,-1}$, $Y_{1,0}$, and $Y_{1,1}$ with the expansion coefficients
$c_{1,-1}=\dfrac{i}{2}$, $c_{1,0}=0$, and $Y_{1,1}= \dfrac{i}{2}$.                   
\section{Scalar product for tensor operators and spherical functions}
\label{appendix B}
The scalar product between two tensor operators $T_A$ and $T_B$ is defined as
\begin{equation}
	\label{Eq.C1}
	\langle{T_{A}|T_{B}}\rangle = \text{tr}(T_{A}^{\dagger}T_{B}),
\end{equation}
where $T_{A}^{\dagger}$ is the adjoint (conjugate transpose) of the operator $T_A$. If the tensor operators $T_A$ and $T_B$ are mapped to spherical droplet functions $f_A(\theta,\phi)$ and $f_B(\theta,\phi)$ using Eq.~\ref{eq.1}, the scalar product of Eq.~\ref{Eq.C1} is by construction identical to the scalar product of the droplet functions defined as~\cite{leiner2017wigner}
\begin{equation}
	\label{Eq.C2}
	 \langle{f_{A}(\theta,\phi)|f_{B}(\theta,\phi)}\rangle = \int_{\theta=0}^{\pi}\int_{\phi=0}^{2\pi}f_{A}^{*}(\theta,\phi) f_{B}(\theta,\phi)\sin(\theta)d\theta d\phi.
\end{equation}
This definition corresponds to the following simple procedure: for each point on the surface of a sphere, the complex conjugate value of the spherical function $f_A$ is multiplied by the value of the spherical function $f_B$ and the integration of the resulting product  $f_A^{*} f_B$ over the surface of the sphere is the value of the desired scalar product $ \langle{f_{A}|f_{B}}\rangle$.

If the spherical functions are only known at a finite number of sample points ($\theta_{i}$,$\phi_{i}$) the scalar product can be approximated by a corresponding discretized scalar product of the form
\begin{equation}
	\label{Eq.C3}
	\langle{f_{A}|f_{B}}\rangle' = \sum_{i} \text{w}_{i} f_{A}^{*}(\theta_{i},\phi_{i})f_{B}(\theta_{i},\phi_{i}) \approx		\langle{f_{A}(\theta,\phi)|f_{B}(\theta,\phi)}\rangle.
\end{equation}
Here, the weights $\text{w}_{i}$  reflect the relative surface area represented by each sample point, which depends on the distribution of the sample points on the sphere for a chosen sampling scheme. \\

As an illustrative example, let us consider the simple case of an equiangular grid, where the sample points are equally distributed along polar ($\theta\in[0,\pi]$) angles as: 
\begin{equation}
	\label{Eq.C41}
	 \theta_{k} = (k-1)d,
\end{equation}
for $k = 1,2,\dots,M+1$. The points are distributed for azimuthal ($\phi\in[0,2\pi)$) angles as:
\begin{equation}
	\label{Eq.C42}
	\phi_{l} = (l-1)d,
\end{equation}
for $l = 1,2,\dots,2M$, where $M$ is a constant and $d = \dfrac{\pi}{M}$ is the angle increment. In this case Eq.~\ref{Eq.C3} can be written in the form  
\begin{equation}
	\label{Eq.C4}
	\langle{f_{A}|f_{B}}\rangle' = \sum_{k}\sum_{l} \text{w}_{k,l} \ f_{A}^{*}(\theta_{k},\phi_{l})f_{B}(\theta_{k},\phi_{l}) \approx		\langle{f_{A}(\theta,\phi)|f_{B}(\theta,\phi)}\rangle .
\end{equation}
At the north pole (where $k=1$, corresponding to $\theta_{1} = 0$) and at the south pole (where $k=M+1$, corresponding to $\theta_{M+1} = \pi$), the weights are
\begin{equation}
	\label{Eq.C5}
	\text{w}_{1,l} = \text{w}_{M+1,l} =  \frac{1}{4M}\Bigg[1-\cos(\frac{d}{2})\Bigg].	
\end{equation}   
For all other sampling points, the weights are given by 
\begin{equation}
	\label{Eq.C6}
	\text{w}_{k,l} = \frac{1}{4M}\Bigg[\cos(\theta_{k}-\frac{d}{2})-\cos(\theta_{k}+\frac{d}{2})\Bigg],	
\end{equation}
where $\theta_{k}$ is given in Eq.~\ref{Eq.C41}. In Eq.~\ref{Eq.C5} and Eq.~\ref{Eq.C6}, $l$ runs from 1 to 2$M$ and in Eq.~\ref{Eq.C6}, $k$ runs from 2 to $M$. \\

In the experiments, for redundancy, for each polar angle $\theta_{k}$, we measured not $2M$ but $2M+1$ phase angles $\phi_{l}$ for sampling points, i.e., according to Eq.~\ref{Eq.C42} the phase angle $\phi_{2M+1} = 2\pi$ and hence corresponds to $\phi_{1} = 0$. This can be simply taken into account by scaling the weights by half for these specific points, i.e., $\text{w}_{k,1} = \text{w}_{k,2M+1} = \frac{1}{2}\text{w}_{k,l}$, where $\text{w}_{k,l}$ is given in Eq.~\ref{Eq.C5} and Eq.~\ref{Eq.C6}.        

\section{Estimation of expectation values}
\label{appendix A}
A general single-qubit state is given by, 
\begin{equation}
	\label{Eq.A1}
	|\psi\rangle = c_{0}|0\rangle+c_{1}|1\rangle,
\end{equation}
where the probability $p_{0} = |c_{0}|^{2}$ to find the state $|0\rangle$ is given by the expectation value $\langle P_{0}\rangle = \langle\psi|P_{0}|\psi\rangle=|c_0|^2$
of the projection operator $P_0= |0\rangle\langle0|$ and the probability $p_{1}=|c_{1}|^{2}$ to find the state in $|1\rangle$ is given by the expectation value $\langle P_{1}\rangle = \langle\psi|P_{1}|\psi\rangle=|c_1|^2$
of the projection operator $P_1= |1\rangle\langle1|$, where the matrix forms of the projection operators $P_{0}$ and $P_{1}$ are
\begin{equation}
	\label{Eq.A2}
	P_{0}= |0\rangle\langle0| = 
	\begin{pmatrix}
		1 & 0 \cr
		0 & 0
	\end{pmatrix}, 
	P_{1}= |1\rangle\langle1| = 
	\begin{pmatrix}
		0 & 0 \cr
		0 & 1
	\end{pmatrix}.
\end{equation}  
As the Pauli operator $\sigma_{z}$ can be written as
\begin{equation}
	\label{Eq.A3}	\sigma_{z} = 
	\begin{pmatrix}
		1 & 0 \cr
		0 & -1
	\end{pmatrix} = P_{0}-P_{1},
\end{equation}
the expectation value of $\sigma_z$ is given by
\begin{equation}
	\label{Eq.A5}
		\langle\sigma_{z}\rangle = \langle P_{0}\rangle - \langle P_{1}\rangle = p_{0}-p_{1}.
\end{equation}
Similarly, the expectation value of the identity matrix $\mathds{1}$ is
\begin{equation}
	\label{Eq.A6}
	\langle\mathds{1}\rangle  = p_{0}+p_{1} = 1.		
\end{equation}
Experimentally, the probabilities $p_{0}$ and $p_{1}$ can be estimated based on the number of shots $N_{0}$ and $N_{1}$ for which the states $|0\rangle$ and $|1\rangle$ are found: $p_{0} \approx N_{0}/N_{s}$ and $p_{1} \approx N_{1}/N_{s}$, where $N_{s} = N_{0}+N_{1}$ is the total number of shots per sampling point.\\

For a two-qubit system ($N=2$), a general quantum state has the form
\begin{equation}
	\label{Eq.A7}
	|\psi\rangle = c_{00}|00\rangle+c_{01}|01\rangle+c_{10}|10\rangle+c_{11}|11\rangle
\end{equation}
and the expectation value of the operators $\mathds{1}^{[2]} = \mathds{1}^{[1]}\otimes\mathds{1}^{[1]}$, $\sigma_{1z}^{[2]} = \sigma_{z}^{[1]}\otimes\mathds{1}^{[1]}$, $\sigma_{2z}^{[2]} = \mathds{1}^{[1]}\otimes\sigma_{z}^{[1]}$, and $(\sigma_{1z}\sigma_{2z})^{[2]}$ are given by
\begin{equation}
	\label{Eq.A8}
	\begin{aligned}
		\langle\mathds{1}^{[2]}\rangle & = \langle P_{00}\rangle + \langle P_{01}\rangle + \langle P_{10}\rangle+\langle P_{11}\rangle = p_{00}+p_{01}+p_{10}+p_{11}\\
		\langle \sigma_{1z}^{[2]}\rangle & = \langle P_{00}\rangle + \langle P_{01}\rangle - \langle P_{10}\rangle-\langle P_{11}\rangle = p_{00}+p_{01}-p_{10}-p_{11}\\
		\langle \sigma_{2z}^{[2]}\rangle & = \langle P_{00}\rangle - \langle P_{01}\rangle + \langle P_{10}\rangle-\langle P_{11}\rangle = p_{00}-p_{01}+p_{10}-p_{11}\\
		\langle(\sigma_{1z}\sigma_{2z})^{[2]}\rangle & = \langle P_{00}\rangle - \langle P_{01}\rangle - \langle P_{10}\rangle+\langle P_{11}\rangle = p_{00}-p_{01}-p_{10}+p_{11},
	\end{aligned}		
\end{equation}
where $P_{ab} = |ab\rangle\langle ab|$ and $p_{ab}$ for $a,b\in\{0,1\}$ are the projection operators and probabilities corresponding to the computational basis states $|ab\rangle$. Hence, the required expectation values for the presented Wigner tomography scheme can be calculated by combining the estimated probabilities $p_{ab}$ to find the system in the computational basis states. Experimentally, the probabilities $p_{00}$, $p_{01}$, $p_{10}$, and $p_{11}$ can be estimated based on the number of shots $N_{00}$, $N_{01}$, $N_{10}$, and $N_{11}$ for which the states $|00\rangle$, $|01\rangle$, $|10\rangle$, and $|11\rangle$ are found: $p_{00}\approx N_{00}/N_{s}$, $p_{01}\approx N_{01}/N_{s}$, $p_{10}\approx N_{10}/N_{s}$, and $p_{11}\approx N_{11}/N_{s}$, where $N_{s} = N_{00}+N_{01}+N_{10}+N_{11}$ is the total number of shots.

Note that other expectation values, such as $\langle \sigma_{1x} \rangle$,  $\langle \sigma_{1x}\sigma_{2x} \rangle$, etc. can be obtained by applying appropriate local unitary operations $u_{n}$ (so-called detection-associated rotation, see Sec.~\ref{Wigner_QST}) before the projective measurement of the computational basis states.   
\section{Generic rotation matrix}
\label{supp:U3}
The state of a qubit can be transformed using unitary rotation operations. The matrix form of the most-general single-qubit rotation, which is also known as $\mathrm{U}_3$ gate, is (up to a global phase) given by 
\begin{equation}
	\label{eq.22a}
	\mathrm{U}_3\mathrm{(}\mathrm{\theta},\mathrm{\phi},\mathrm{\lambda})= 
	\begin{pmatrix} 
		\text{cos}(\theta/2) 
		& 
		-\text{e}^{i\lambda}\text{sin}(\theta/2) \cr
		\text{e}^{i\phi}\text{sin}(\theta/2) & \text{e}^{i(\lambda+\phi)}\text{cos}(\theta/2)
	\end{pmatrix}.
\end{equation}
This can also be written in the form of the following Euler angle decomposition
\begin{equation}
	\label{eq.22b}
	\begin{split}
		\mathrm{U}_3\mathrm{(}\mathrm{\theta},\mathrm{\phi},\mathrm{\lambda}) & = \textit{RZ}(\mathrm{\phi})\textit{RY}(\mathrm{\theta})\textit{RZ}(\mathrm{\lambda})\\
		&=\textit{RZ}(\mathrm{\phi})\textit{RX}(-\pi/2)\textit{RZ}(\mathrm{\theta})\textit{RX}(\pi/2)\textit{RZ}(\mathrm{\lambda}).
	\end{split}
\end{equation}
This corresponds to a rotation of angle $\lambda$ around the $z$ axis, followed by a rotation $\theta$ around the $y$ axis, and followed by a rotation of $\phi$ around the $z$ axis. The values of the three Euler angles $\theta,\phi,\lambda$ can be adjusted to implement any desired unitary operation or a gate~\cite{goldstein2002classical}. For example, $\mathrm{U}_3 (\theta,-\pi/2,\pi/2)$ corresponds to a rotation around the $x$ axis by an angle $\theta$ and $\mathrm{U}_3 (\theta,0,0)$ corresponds to a rotation around the $y$ axis by an angle $\theta$. The rotation operation $R_{\alpha\beta}$ used for scanning in the state tomography algorithm corresponds in this nomenclature to $\mathrm{U}_3 (\beta,\alpha,0)$ for $\beta\in[0,\pi]$ and $\alpha\in[0,2\pi]$. Similarly, different steps of the presented Wigner tomography algorithm can be designed with the help of the general rotation gate $\mathrm{U}_3$.

\section{Quantum circuit and plots of corresponding expectation values for one-qubit Wigner state tomography}
\label{Sec.:Qcircuits}
\begin{figure}[h]
	\centering
	\begin{minipage}{.48\textwidth}
		\includegraphics[scale=1.1]{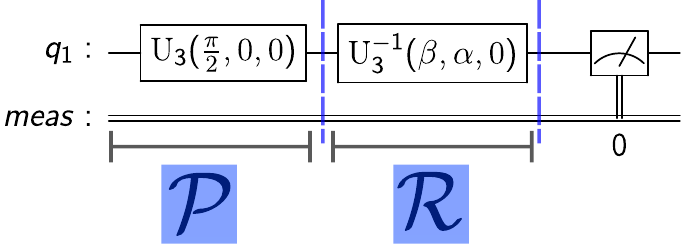}
		\caption{Quantum circuit for performing Wigner state tomography for the example of the state $|\psi\rangle=\frac{|0\rangle+|1\rangle}{\sqrt{2}}$. The first and second block of the circuit corresponds to the Preparation ($\mathcal{P}$) and Rotation ($\mathcal{R}$) step of state tomography from left to right. The superposition state is prepared from the initial state $|\psi\rangle_i=|0\rangle$. The $\mathrm{U}_{3}$ gate used in the circuit is discussed in supplementary Sec.~\ref{supp:U3}.}
		\label{fig:QST1_1}
	\end{minipage}%
	\hfill
	\begin{minipage}{.48\textwidth}
		\centering
		\includegraphics[scale=0.8]{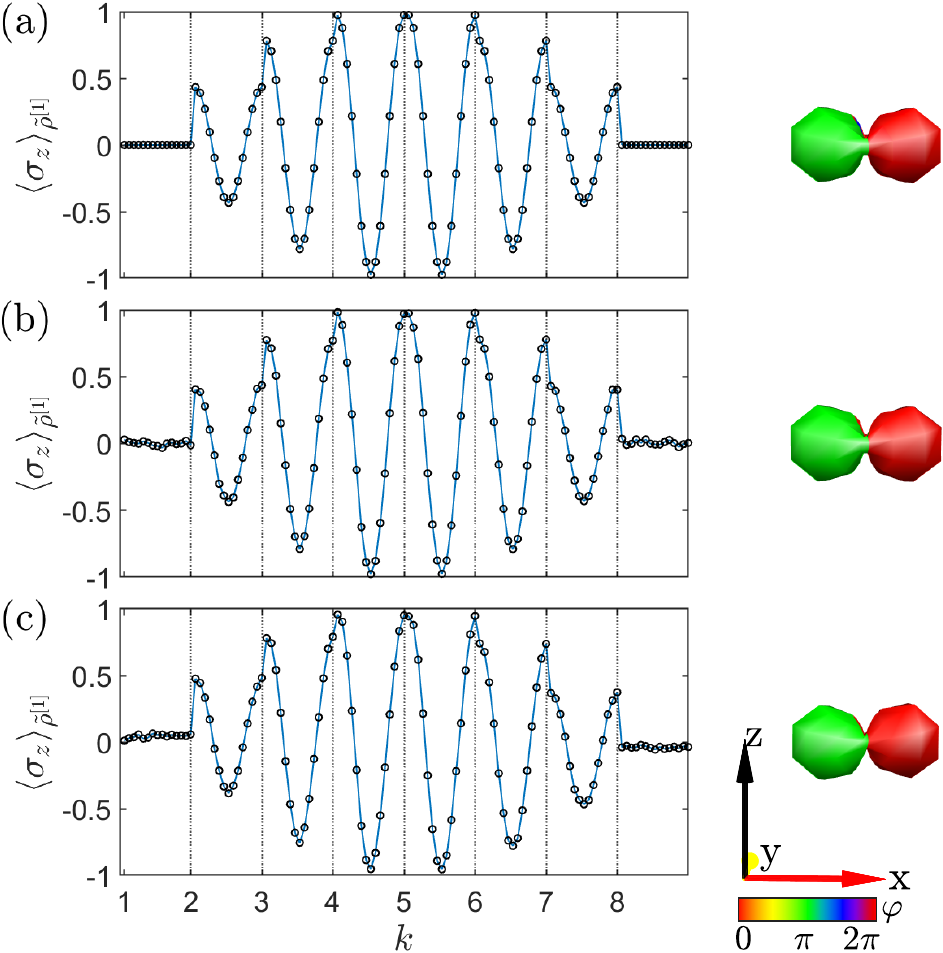}
		\caption{The expectation values and the droplet corresponding to the rank $j=1$ component of a quantum state $|\psi\rangle=\frac{1}{\sqrt{2}}(|0\rangle+|1\rangle)$: (a) ideal simulation, (b) simulation with shot noise, and (c) experimental data. The expectation values are calculated for the combinations of eight polar angles $\beta_k = (k-1)\dfrac{\pi}{7}$ and fifteen azimuthal angles $\alpha_{l}= (l-1)\dfrac{\pi}{7}$. In the figure, for each value of $k$ the azimuthal counter ${l}$ is incremented from 1 to 15.}
		\label{fig:expec&drops}
	\end{minipage}
\end{figure}

In Fig.~\ref{fig:QST1_1}, we show the quantum circuit for performing state tomography of a single-qubit considering the example shown in Fig.~\ref{fig:QST result single qubit}b in the main text, where the state of the qubit is the superposition state $|\psi\rangle = \frac{1}{\sqrt{2}}(|0\rangle+|1\rangle)$. As shown in Fig.~\ref{fig:QST1_1}, this state is prepared from the initial state $|\psi_{i}\rangle=|0\rangle$ using $\mathrm{U}_3 (\frac{\pi}{2},0,0)$ (alternately, a Hadamard gate could be used). This preparation step is followed by the rotation step $\mathcal{R}$ which is implemented by $\mathrm{U}_3^{-1} (\beta,\alpha,0) = \mathrm{U}_3 (-\beta,0,-\alpha)$.\\

In Fig.~\ref{fig:expec&drops}, we plot corresponding simulated and experimental expectation values $\langle\sigma_{z}\rangle_{\tilde{\rho}^{[1]}}$ for all combinations of scanning angles $\beta_k = (k-1)\dfrac{\pi}{7}$ and $\alpha_{l}= (l-1)\dfrac{\pi}{7}$, where $k = 1,2,\dots,8$ and $l = 1,2,\dots,15$. Fig.~\ref{fig:expec&drops}a shows the ideal expectation values with no noise, Fig.~\ref{fig:expec&drops}b shows the simulated expectation values with shot noise for number of shots $N_{s} = 8192$, and Fig.~\ref{fig:expec&drops}c shows the experimental expectation values for the same number of shots $N_{s}$. Based on these expectation values, the rank $j=1$ droplet function $f_1^{(1)}$ can be calculated using Eq.~\ref{eq.20} and Fig.~\ref{fig:expec&drops} shows also plots of the corresponding tomographed droplets.
\section{Generating basis droplets}
\label{Sec.:Sph_basis}
Here, we summarize the explicit form of the ideal spherical droplet functions $	f_{\sigma_{k}}(\beta,\alpha)$ in terms of linear combinations of complex spherical harmonics $Y_{jm}(\beta,\alpha)$. This decomposition is based on the general basis transformations from Pauli to tensor operators, c.f. Ref.~\cite{DROPS_main}. As shown in Sec.~\ref{Sec.:one qubit QST} and Sec.~\ref{Sec.:two qubit QST}, these basis droplet functions $	f_{\sigma_{k}}(\beta,\alpha)$ can be used to estimate the density matrix corresponding to an experimentally measured droplet.

\subsection{One qubit}
\label{Sec.:Basis_single}
The basis droplet functions ($f_{\sigma_{k}}$) for $k\in\{0,1,2,3\}$ in terms of complex spherical harmonics ($Y_{jm}$) for one qubit are:
\begin{equation}
	\label{Eq.basis_single}
	\begin{aligned}
		f_{\sigma_{0}}(\beta,\alpha) & = \sqrt{2}Y_{00}(\beta,\alpha)\\
		f_{\sigma_{1}} = f_{\sigma_{x}}(\beta,\alpha) & = Y_{1-1}(\beta,\alpha)-Y_{11}(\beta,\alpha) \\
		f_{\sigma_{2}} = f_{\sigma_{y}}(\beta,\alpha) & = i(Y_{1-1}(\beta,\alpha)+Y_{11}(\beta,\alpha)) \\
		f_{\sigma_{3}} = f_{\sigma_{z}}(\beta,\alpha) & = \sqrt{2}Y_{10}(\beta,\alpha),
	\end{aligned}		
\end{equation}
where $\sigma_{0}$ corresponds to the $2\times2$ identity operator. 
\subsection{Two qubits}
\label{Sec.:Basis_two}
The basis droplet functions ($f_{\sigma_{k}}$) for $k\in\{0,1,\dots,15\}$ in terms of complex spherical harmonics ($Y_{jm}$) for two qubit are:
\begin{equation}
	\label{Eq.basis_single}
	\begin{aligned}
		f_{\sigma_{0}}(\beta,\alpha) & = Y_{00}(\beta,\alpha)\\
		f_{\sigma_{1}} = f_{\sigma_{1x}}(\beta,\alpha) & = \frac{1}{\sqrt{2}} (Y_{1-1}(\beta,\alpha)-Y_{11}(\beta,\alpha)) \\
		f_{\sigma_{2}} = f_{\sigma_{1y}}(\beta,\alpha) & = \frac{i}{\sqrt{2}} (Y_{1-1}(\beta,\alpha)+Y_{11}(\beta,\alpha)) \\
		f_{\sigma_{3}} = f_{\sigma_{1z}}(\beta,\alpha) & = Y_{10}(\beta,\alpha)\\
		f_{\sigma_{4}} = f_{\sigma_{2x}}(\beta,\alpha) & = \frac{1}{\sqrt{2}} (Y_{1-1}(\beta,\alpha)-Y_{11}(\beta,\alpha)) \\
		f_{\sigma_{5}} = f_{\sigma_{2y}}(\beta,\alpha) & = \frac{i}{\sqrt{2}} (Y_{1-1}(\beta,\alpha)+Y_{11}(\beta,\alpha)) \\
		f_{\sigma_{6}} =f_{\sigma_{2z}}(\beta,\alpha) & = Y_{10}(\beta,\alpha)\\
		f_{\sigma_{7}} =f_{\sigma_{1x2x}}(\beta,\alpha) & = \frac{1}{\sqrt{3}} Y_{00}(\beta,\alpha)+\frac{1}{2} Y_{2-2}(\beta,\alpha)-\frac{1}{\sqrt{6}} Y_{20}(\beta,\alpha)+\frac{1}{2} Y_{22}(\beta,\alpha) \\
		f_{\sigma_{8}} =f_{\sigma_{1x2y}}(\beta,\alpha) & = \frac{1}{\sqrt{2}} Y_{10}(\beta,\alpha)+\frac{i}{2}Y_{2-2}(\beta,\alpha)-\frac{i}{2}Y_{22}(\beta,\alpha) \\
		f_{\sigma_{9}} =f_{\sigma_{1x2z}}(\beta,\alpha) & = -\frac{i}{2}Y_{1-1}(\beta,\alpha)-\frac{i}{2}Y_{11}(\beta,\alpha)+\frac{1}{2}Y_{2-1}(\beta,\alpha)-\frac{1}{2}Y_{21}(\beta,\alpha)\\
		f_{\sigma_{10}} =f_{\sigma_{1y2x}}(\beta,\alpha) & = -\frac{1}{\sqrt{2}} Y_{10}(\beta,\alpha)+\frac{1}{2} Y_{2-2}(\beta,\alpha)-\frac{i}{2} Y_{22}(\beta,\alpha)\\
		f_{\sigma_{11}} =f_{\sigma_{1y2y}}(\beta,\alpha) & = \frac{1}{\sqrt{3}} Y_{00}(\beta,\alpha)-\frac{1}{2} Y_{2-2}(\beta,\alpha)-\frac{1}{\sqrt{6}} Y_{20}(\beta,\alpha)-\frac{1}{2} Y_{22}(\beta,\alpha)\\
		f_{\sigma_{12}} =f_{\sigma_{1y2z}}(\beta,\alpha) & = \frac{1}{2} Y_{1-1}(\beta,\alpha)-\frac{1}{2} Y_{11}(\beta,\alpha)+\frac{i}{2} Y_{2-1}(\beta,\alpha)+\frac{i}{2} Y_{21}(\beta,\alpha)\\
		f_{\sigma_{13}} =f_{\sigma_{1z2x}}(\beta,\alpha) & = \frac{i}{2} Y_{1-1}(\beta,\alpha)+\frac{i}{2} Y_{11}(\beta,\alpha)+\frac{1}{2} Y_{2-1}(\beta,\alpha)-\frac{1}{2} Y_{21}(\beta,\alpha)\\
		f_{\sigma_{14}} =f_{\sigma_{1z2y}}(\beta,\alpha) & = -\frac{1}{2} Y_{1-1}(\beta,\alpha)+\frac{1}{2} Y_{11}(\beta,\alpha)+\frac{i}{2} Y_{2-1}(\beta,\alpha)+\frac{i}{2} Y_{21}(\beta,\alpha)\\
		f_{\sigma_{15}} =f_{\sigma_{1z2z}}(\beta,\alpha) & = \frac{1}{\sqrt{3}} Y_{00}(\beta,\alpha)+\sqrt{\frac{2}{3}} Y_{20}(\beta,\alpha),
	\end{aligned}		
\end{equation}
where $\sigma_{0}$ corresponds to the $4\times4$ identity operator.

\section{Temporal averaging to create maximally mixed states}
\label{sec.temporal_avg}  
For Wigner quantum process tomography, the preparation of the system qubit in a \textit{maximally} mixed state is required (see preparation ($\mathcal{P}$) step in Sec.~\ref{sec.WQPT}). As stated in Sec~\ref{WQPT_expt_single_qubit}, in our experimental implementation, a temporal averaging approach~\cite{knill_temporal,preskill1998lecture} was used to prepare the maximally mixed state. A detailed explanation of this approach and alternative methods are provided in the following.
\subsection{Creating maximally mixed state from pure states}
In general, there is an infinite number of different ways to prepare the ``maximally mixed state" $\rho_{mm}$ of a single-qubit ($N=1$) based on an ensemble of pure single-qubit states~\cite{preskill1998lecture}. For example, the maximally mixed state can be realized using an ensemble of two extremal pure states $|{\uparrow_{\hat{n}}}\rangle$ and $|{\downarrow_{\hat{n}}}\rangle$ corresponding to two antipodal points on the Bloch sphere:
\begin{equation}
	\label{Eq.D4}
	\rho_{mm}^{[1]} = \frac{1}{2}\mathds{1}^{[1]} = \frac{1}           {2}\begin{pmatrix}
	1 & 0 \cr
	0 & 1\end{pmatrix} = \frac{1}{2} |{\uparrow_{\hat{n}}}\rangle \langle{\uparrow_{\hat{n}}}| + \frac{1}{2} |{\downarrow_{\hat{n}}}\rangle \langle{\downarrow_{\hat{n}}}|.
\end{equation}
Hence, for the single-qubit case, the preparation of a maximally mixed state $\rho_{mm}^{[1]}$ (see Eq.~\ref{Eq.D4}) can be realized by simply repeating the experiment twice and averaging the measurement results (which we refer to as \textit{temporal} averaging here): once by starting with the pure-state $|0\rangle$ (corresponding to the density operator $|0\rangle\langle0| = \bigl( \begin{smallmatrix} 1 & 0\\0 & 0 \end{smallmatrix} \bigr)$) and once by starting with the pure-state $|1\rangle$ (corresponding to the density operator $|1\rangle\langle1| = \bigl( \begin{smallmatrix} 0 & 0\\0 & 1 \end{smallmatrix} \bigr)$). Note that the average density operator of the two experiments is identical to the desired maximally mixed state $\rho_{mm}^{[1]}$.\\

An alternative way to create a maximally mixed state is by creating a maximal entanglement between the system and an ancilla qubit and partially tracing out of the ancilla qubit. However, this approach requires additional resources (both in a number of qubits and gates) compared to the scheme presented here.

\subsection{Creating the maximally mixed state of two-qubit by temporal averaging}
For the case of a two-qubit ($N = 2$) system, the maximally mixed state $\rho_{mm}^{[2]}$ is of the form
\begin{equation}
	\label{Eq.D41}
	\rho_{mm}^{[2]} = \frac{1}{4}\mathds{1}^{[2]} = \frac{1}{4}\begin{pmatrix}
	1 & 0 & 0 & 0\cr
	0 & 1 & 0 & 0\cr
	0 & 0 & 1 & 0\cr
	0 & 0 & 0 & 1\cr\end{pmatrix} =  \frac{1}{4}|00\rangle\langle{00}|+\frac{1}{4}|01\rangle\langle{01}|+\frac{1}{4}|10\rangle\langle{10}|+\frac{1}{4}|11\rangle\langle{11}|.
\end{equation}
In this case, the preparation of $\rho_{mm}^{[2]}$ can be realized by simply repeating the experiment four times (for all the computational basis states) and averaging the measurement results.
\subsection{Creating the state $|+\rangle\langle+| \otimes \frac{1}{2^{N}}\mathds{1}^{[N]}$ required for process tomography of unitary gates}
Here we give a detailed calculation of the preparation step $\mathcal{{P}}$ (see Sec.~\ref{WQPT}) for one and two-qubit Wigner process tomography. For single-qubit process tomography, two qubits $q_{0}$ and $q_{1}$ are required and the preparation step requires preparing qubit $q_{0}$ in an equal superposition state $|+\rangle = \dfrac{1}{\sqrt{2}}(|0\rangle+|1\rangle)$ and $q_{1}$ in a maximally mixed state $\dfrac{1}{2}\mathds{1}^{[1]}$. The corresponding density matrix is given by
\begin{equation}
	\label{Eq.D42}
	\rho_{0}^{[2]} = |+\rangle\langle+| \otimes \frac{1}{2}\mathds{1}^{[1]} = \frac{1}{2}\begin{pmatrix}
		1 & 1 \cr
		1 & 1\end{pmatrix} \otimes \frac{1}{2}\begin{pmatrix}
		1 & 0 \cr
		0 & 1\end{pmatrix} = \frac{1}{4}\begin{pmatrix}
		1 & 0 & 1 & 0\cr
		0 & 1 & 0 & 1\cr
		1 & 0 & 1 & 0\cr
		0 & 1 & 0 & 1\cr\end{pmatrix},	
\end{equation}
This maximally mixed state of qubit $q_{1}$ is achieved using deterministic temporal averaging (see Eq.~\ref{Eq.D4}), i.e., by averaging measurement result of two experiments initialized in states $|\psi_{1}\rangle = |+\rangle\otimes|0\rangle = |+0\rangle$ and $|\psi_{2}\rangle = |+\rangle\otimes|1\rangle = |+1\rangle$. The two corresponding density matrices are
\begin{equation}
	\label{Eq.D43}
	\rho_{1}^{[2]} = |\psi_{1}\rangle\langle\psi_{1}| = |{+0}\rangle\langle{+0}| = \frac{1}{2} \begin{pmatrix}
	1 & 0 & 1 & 0\cr
	0 & 0 & 0 & 0\cr
	1 & 0 & 1 & 0\cr
	0 & 0 & 0 & 0\cr\end{pmatrix},	
\end{equation}
and 
\begin{equation}
	\label{Eq.D44}
	\rho_{2}^{[2]} = |\psi_{2}\rangle\langle\psi_{2}| = |{+1}\rangle\langle{+1}| = \frac{1}{2} \begin{pmatrix}
		0 & 0 & 0 & 0\cr
		0 & 1 & 0 & 1\cr
		0 & 0 & 0 & 0\cr
		0 & 1 & 0 & 1\cr\end{pmatrix}.	
\end{equation} 
Therefore, the density matrix $\rho_{0}^{[2]}$ can be prepared by averaging the density operators $\rho_{1}^{[2]}$ and $\rho_{2}^{[2]}$:
\begin{equation}
	\label{Eq.D45}
	\rho_{0}^{[2]} = \frac{1}{2}\rho_{1}^{[2]} + \frac{1}{2}\rho_{2}^{[2]}.
\end{equation}
Similarly, for a two-qubit process tomography, three qubits $q_{0}$, $q_{1}$, and $q_{2}$ are required. The preparation step requires preparing qubit $q_{0}$ in an equal superposition state $|+\rangle$ and qubits $q_{1}$ and $q_{2}$ in a maximally mixed state $\dfrac{1}{4}\mathds{1}^{[2]}$. This resultant density matrix $\rho_{0}^{[3]}$ can be achieved using Eq.~\ref{Eq.D41} as follows:
\begin{equation}
	\label{Eq.D46}
	\rho_{0}^{[3]} = |+\rangle\langle+| \otimes \frac{1}{4}\mathds{1}^{[2]} = \frac{1}{8}|{+00}\rangle\langle{+00}|+\frac{1}{8}|{+01}\rangle\langle{+01}|+\frac{1}{8}|{+10}\rangle\langle{+10}|+\frac{1}{8}|{+11}\rangle\langle{+11}|.
\end{equation}
This corresponds to performing four different experiments and averaging the output density operators.

\subsection{Expectation values for mixed states}
Here, we explicitly show that the expectation values for a mixed state can be expressed as an average of the expectation values of pure states. The expectation value of an observable $A$ is given by
\begin{equation}
	\label{Eq.D1}
	\langle{A}\rangle_{\rho} = \text{tr}(A\rho). 
\end{equation}
The density matrix $\rho$ can be written in its spectral decomposition as
\begin{equation}
	\label{Eq.D2}
	\rho = \sum_{k=1}^{2^{N}}p_{k}|\beta_k\rangle\langle{\beta_k}|,
\end{equation} 
where $p_k$ are non-negative coefficients which add up to one and $|\beta_k\rangle$ are the basis states. Eq.~\ref{Eq.D1} can be rewritten as:
\begin{equation}
	\label{Eq.D3}
	\begin{split}
		\langle{A}\rangle_{\rho} &= \text{tr}\bigg(A\sum_{k=1}^{2^{N}}p_{k}|\beta_k\rangle\langle{\beta_k}|\bigg)\\
		&=  \sum_{k=1}^{2^N}p_{k}\langle{A}\rangle_{|\beta_k\rangle\langle{\beta_k}|}.
	\end{split}
\end{equation} 

Hence, for the special case where $\rho$ is the completely mixed state with identical coefficients $p_{k} = \frac{1}{2^{N}}$, the expectation value $\langle{A}\rangle_{\rho}$ is simply the average of the expectation values for the pure states $|\beta_k\rangle$. For example, for a single qubit the maximally mixed state $\mathds{1}^{[1]}/2$ can be written as:
where $|\beta_1\rangle = |0\rangle$, $|\beta_2\rangle = |1\rangle$, and $p_{1}=p_{2}=\frac{1}{2}$. This implies that experimentally estimating the expectation value of an observable $A$ for the mixed state $\frac{\mathds{1}^{[1]}}{2}$ can be realized by averaging the expectation values from two different experiments, one in which the qubit is in the pure state $|0\rangle$, and another in which the qubit is in the pure state $|1\rangle$. 
\section{Additional figures for Wigner state and process tomography results}
\label{Sec.:Add_result_figs} 
\begin{figure}[h]
	\centering
	\includegraphics[scale=0.65]{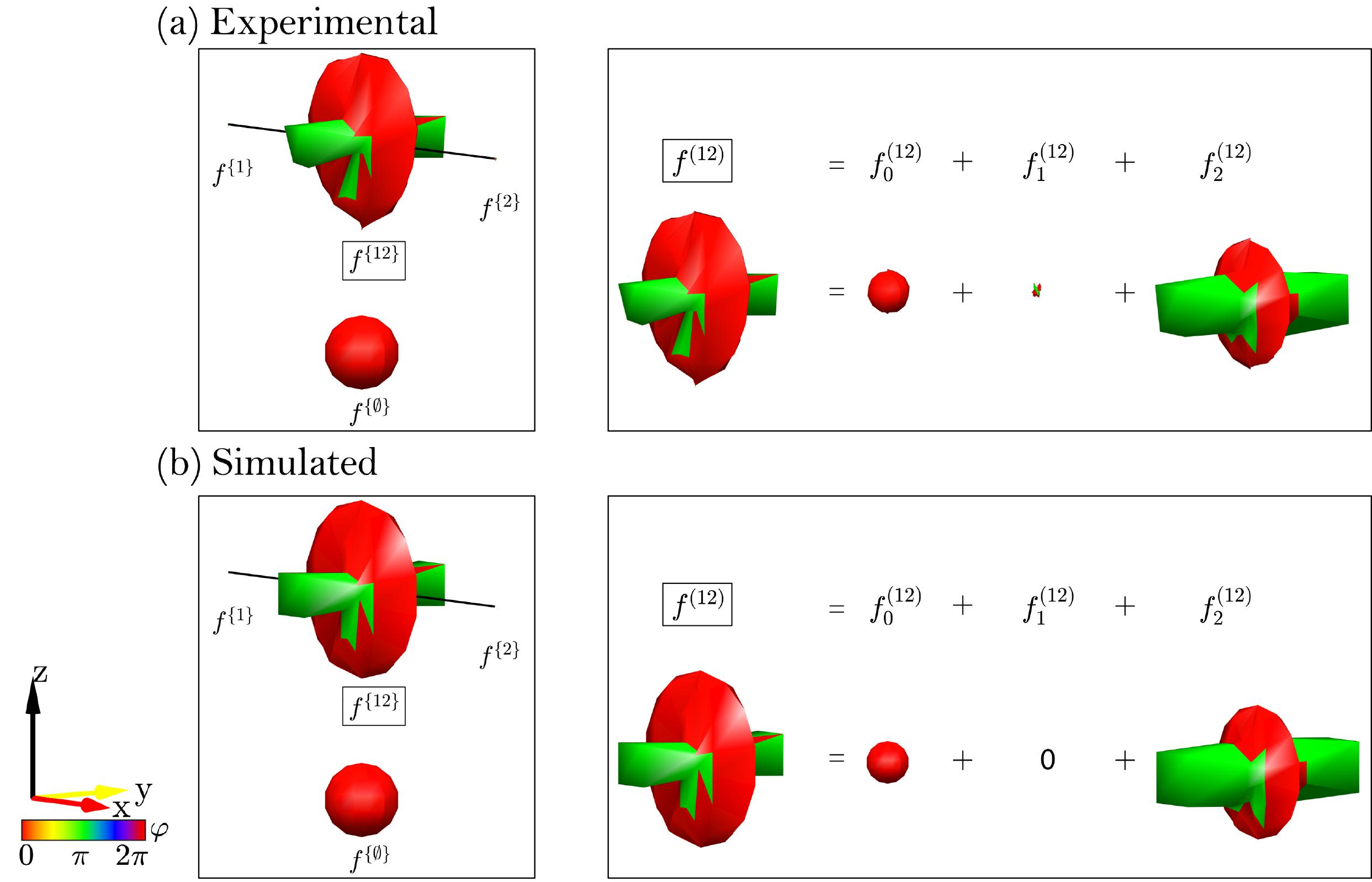}
	\caption{DROPS representation of the Bell state $|\psi\rangle = \frac{1}{\sqrt{2}}(|00\rangle+|11\rangle)$. Experimentally tomographed droplets are shown in the upper panel (a), whereas the simulated droplets are shown in the lower panel (b). The right panels show the respective bilinear droplet function $f^{(12)}$ (box) decomposed into its multipole contribution $f_{j}^{(12)}$ with $j\in\{0,1,2\}$. Here, both experimental and simulated droplets are plotted with same resolution.}
	\label{fig:Bell_result_ranks}
\end{figure}

\begin{figure}[h]
	\centering
	\includegraphics[scale=0.65]{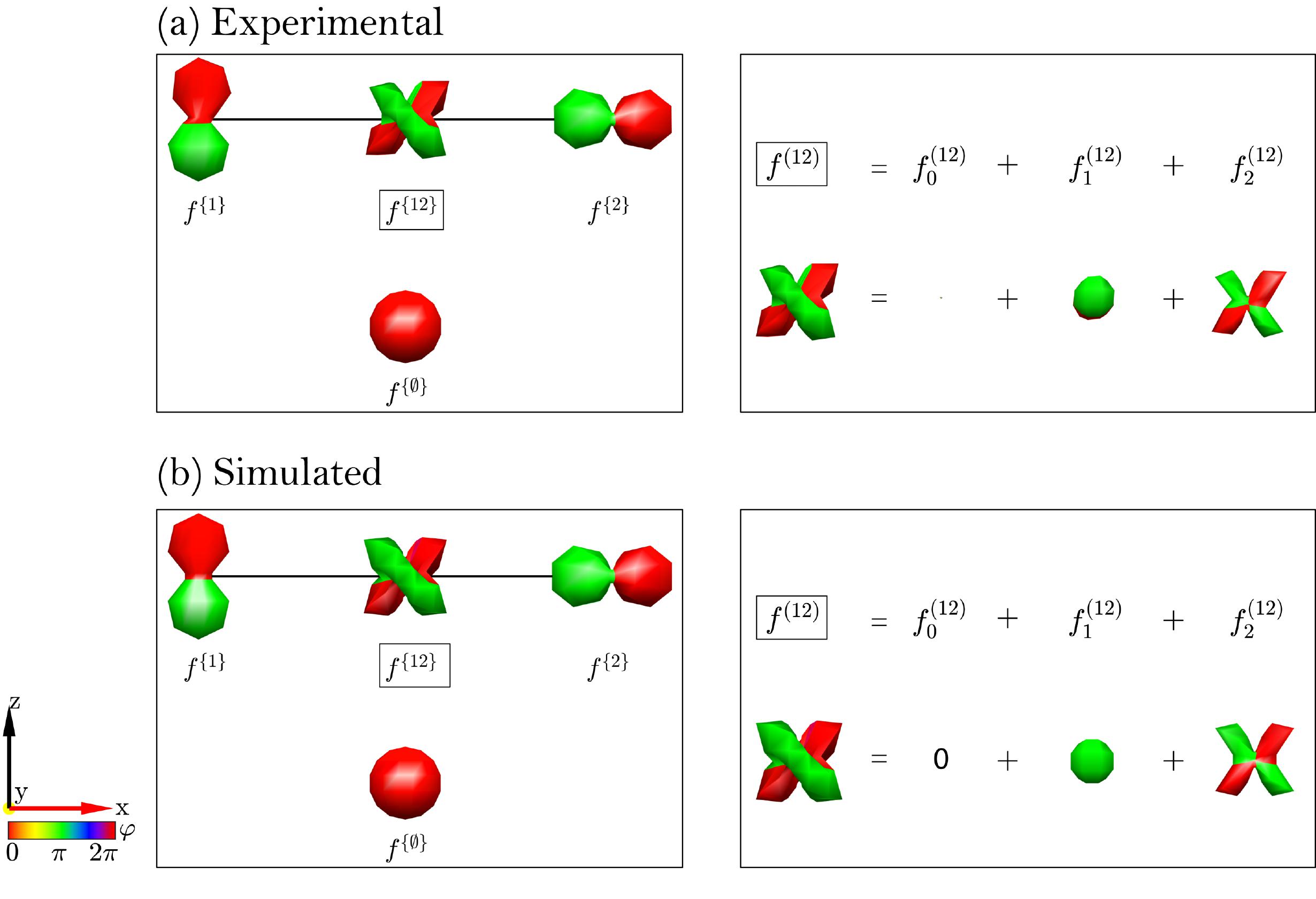}
	\caption{DROPS representation of the state $|\psi\rangle = \frac{1}{\sqrt{2}}(|00\rangle+|01\rangle)$. Experimentally tomographed droplets are shown in the upper panel (a), whereas the simulated droplets are shown in the lower panel (b). Right panels show the respective bilinear droplet function $f^{(12)}$ (box) decomposed into its multipole contribution $f_{j}^{(12)}$ with $j\in\{0,1,2\}$. Here, both experimental and simulated droplets are plotted with the same resolution.}
	\label{fig:Sep_result}
\end{figure}

\begin{figure}[h]
	\centering
	\includegraphics[scale=1.1]{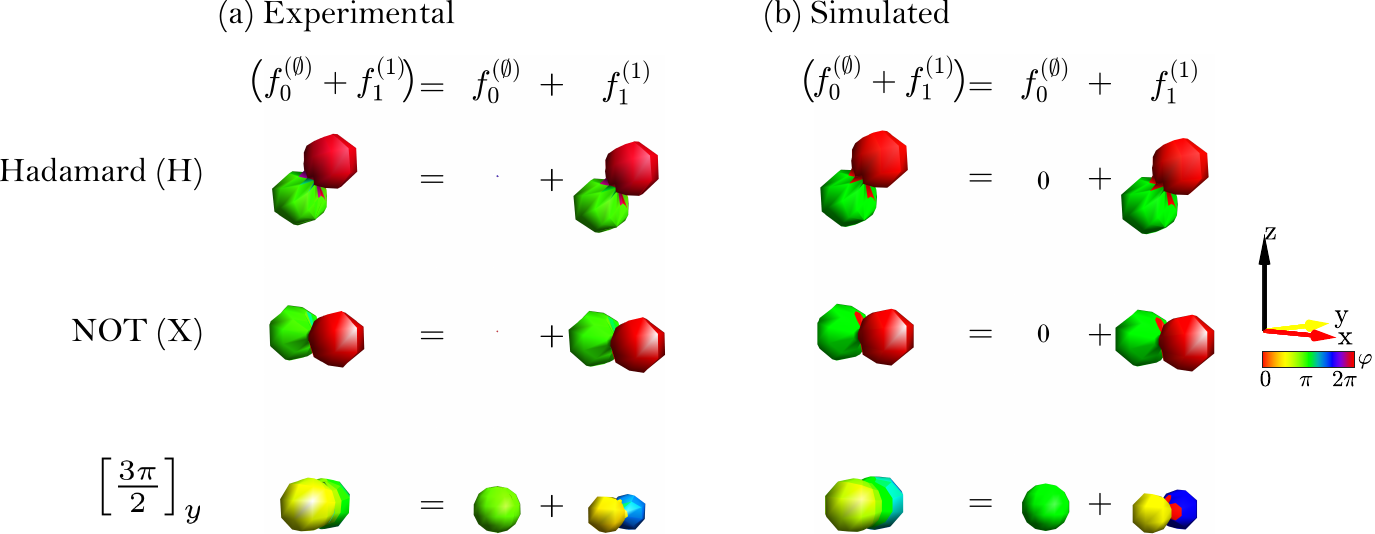}
	\caption{Experimentally tomographed (a) and simulated (b) droplets of different quantum processes. The rank $j=0$ droplet $f_{0}^{(\emptyset)}$ and $j=1$ droplet $f_{1}^{(1)}$ are shown here. Both experimental and simulated droplets are plotted with the same resolution.}
	\label{fig:QPT_results_ranks}
\end{figure}

Whereas in Figs.~\ref{fig:BellNew} and \ref{fig:00+01New} of the main text all bilinear terms $f_{j}^{\{12\}}$ with $j\in\{0,1,2\}$ are merged into a single droplet $f^{\{12\}}$, for completeness, here we also plot the individual droplets $f_{j}^{\{12\}}$ separately in Figs.~\ref{fig:Bell_result_ranks} and~\ref{fig:Sep_result}. Furthermore, in contrast to Figs.~\ref{fig:BellNew} and \ref{fig:00+01New}, where the simulated droplets were calculated and plotted with high resolution using a large number of sampling points, in Figs.~\ref{fig:Bell_result_ranks} and~\ref{fig:Sep_result} we plot the simulated droplet with the same resolution used in the experiments, i.e., eight polar $\beta\in\{0,\frac{\pi}{7},\cdots\pi\}$ and fifteen azimuthal $\alpha\in\{0,\frac{2\pi}{14},\cdots2\pi\}$ angles. For the same resolution, simulated and experimental droplet functions show comparable plotting artifacts of the Matlab display function. Similarly, Fig.~\ref{fig:QPT_results_ranks} shows the individual droplets $f_{0}^{(\emptyset)}$ and $f_{1}^{(1)}$ for the process tomography results shown in Fig.~\ref{fig:QPT_results} of the main text.     

\section{Proof of equivalence of Eq.~\ref{eq.38} and Eq.~\ref{eq.39}}
\label{WQPT_supp. sec.}
After the mapping of the process matrix onto the density matrix, the Eq.~\ref{eq.34} can be written as:
\begin{equation}
	\label{eq.wqpt_1}
	\rho_{U}^{[N+1]} = \frac{1}{2^{N+1}}(\sigma^{-}\otimes U^{[N]}+\sigma^{+}\otimes (U^{[N]})^{\dagger}+\mathds{1}\otimes\mathds{1}^{[N]}),
\end{equation} 
where  $\sigma^{+} = \frac{1}{2}(\sigma_{x}+i\sigma_{y})$, and $\sigma^{-} = \frac{1}{2}(\sigma_{x}-i\sigma_{y})$. The equivalence of Eq.~\ref{eq.39} and Eq.~\ref{eq.38} can be shown by the following steps which transform Eq.~\ref{eq.39} to Eq.~\ref{eq.38}: 
\begin{equation}
	\label{eq.wqpt_2}
	\begin{aligned}
		s_{j}\langle{\sigma^{+}\otimes T_{j,\alpha\beta}^{(\ell)[N]}}\rangle_{\rho_{U}^{[N+1]}} &= s_{j}\text{tr}[(\sigma^{+}\otimes T_{j,\alpha\beta}^{(\ell)[N]})\rho_{U}^{[N+1]}]\\
		& = s_{j}\text{tr}\Bigg[\begin{pmatrix}
			0^{[N]} & T_{j,\alpha\beta}^{(\ell)[N]} \cr
			0^{[N]} & 0^{[N]}
		\end{pmatrix}
		\begin{pmatrix}
			\mathds{1}^{[N]} 
			& 
			(U^{[N]})^{\dagger}  \cr
			U^{[N]}  & \mathds{1}^{[N]} 
		\end{pmatrix} \Bigg]\\
		& = s_{j}\text{tr}\Bigg[\begin{pmatrix}
			T_{j,\alpha\beta}^{(\ell)[N]} U^{[N]} & T_{j,\alpha\beta}^{(\ell)[N]} \cr
			0^{[N]} & 0^{[N]}
		\end{pmatrix}\Bigg] \\
		& = s_{j}\text{tr}[T_{j,\alpha\beta}^{(\ell)[N]} U^{[N]}]\\
		& = s_{j}\text{tr}[(T_{j,\alpha\beta}^{(\ell)[N]})^{\dagger} U^{[N]}]\\
		& = s_{j}\langle{T_{j,\alpha\beta}^{(\ell)[N]}}|{U^{[N]}}\rangle.
	\end{aligned}
\end{equation}
Note that axial tensor operators are Hermitian, i.e., $({T_{j,\alpha\beta}^{(\ell)[N]}})^\dagger = ({T_{j,\alpha\beta}^{(\ell)[N]}})$~\cite{leiner2017wigner}.
\section{Plot of the mean fidelity as a function of the total number of shots}
\label{Sec.:shots_study}
\begin{figure}[h]
	\centering
	\includegraphics[scale=1]{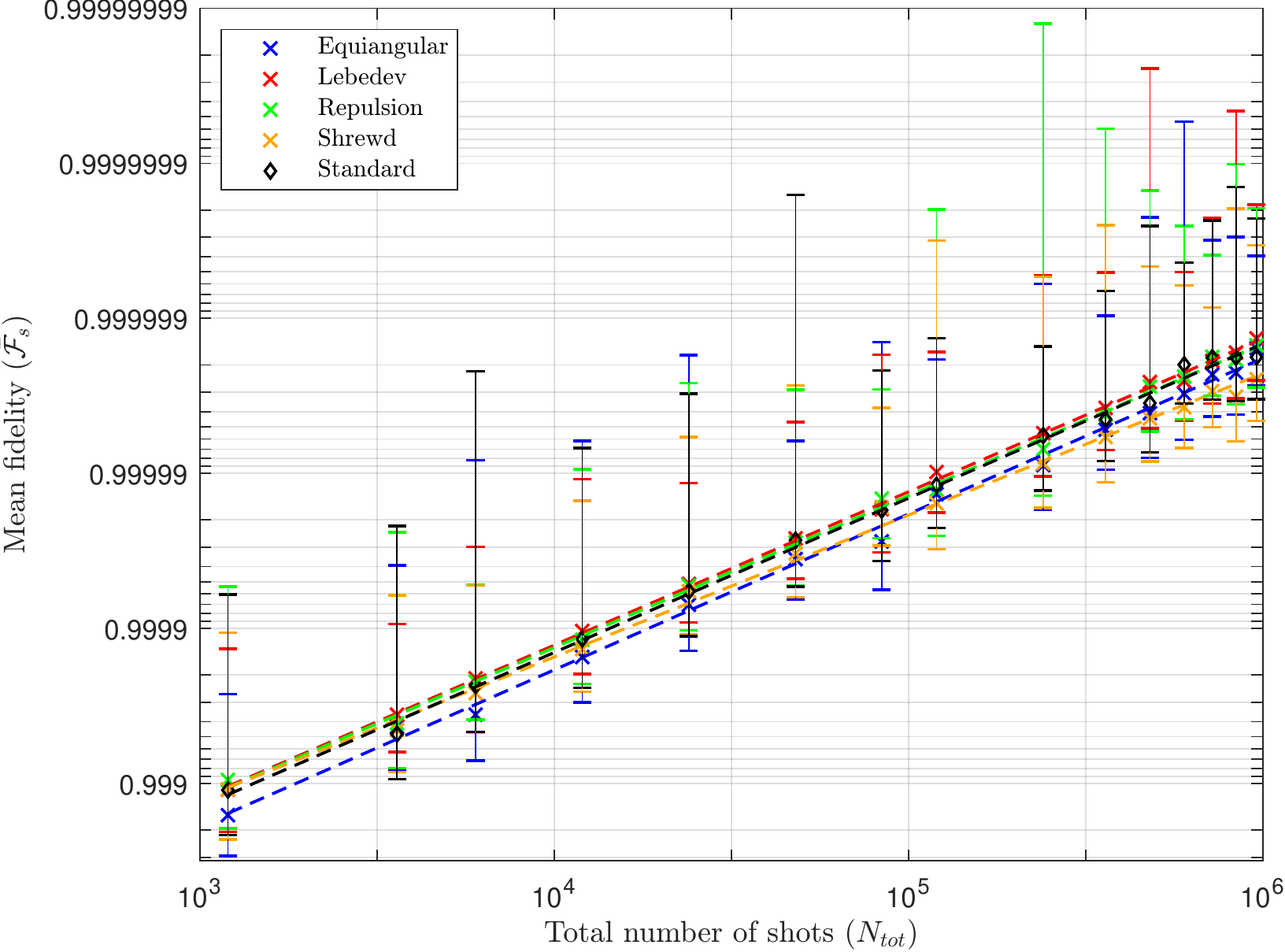}
	\caption{Plot of the mean fidelity ($\bar{\mathcal{F}_s}$) as a function of the total number of shots ($N_{tot}$) for different sampling techniques and for the standard state tomography method applied to the quantum state $|\psi\rangle = (-0.69-0.098i)|0\rangle+(0.66+0.30i)|1\rangle$. The mean fidelity is calculated by repeating the simulation 100 times for each data point. In the simulation, only the noise due to a limited number of shots is considered. The standard deviations are shown by the vertical bars for each point.}
	\label{fig:ShotSim_wSD}
\end{figure}

\begin{figure}[h]
	\centering
	\includegraphics[scale=1]{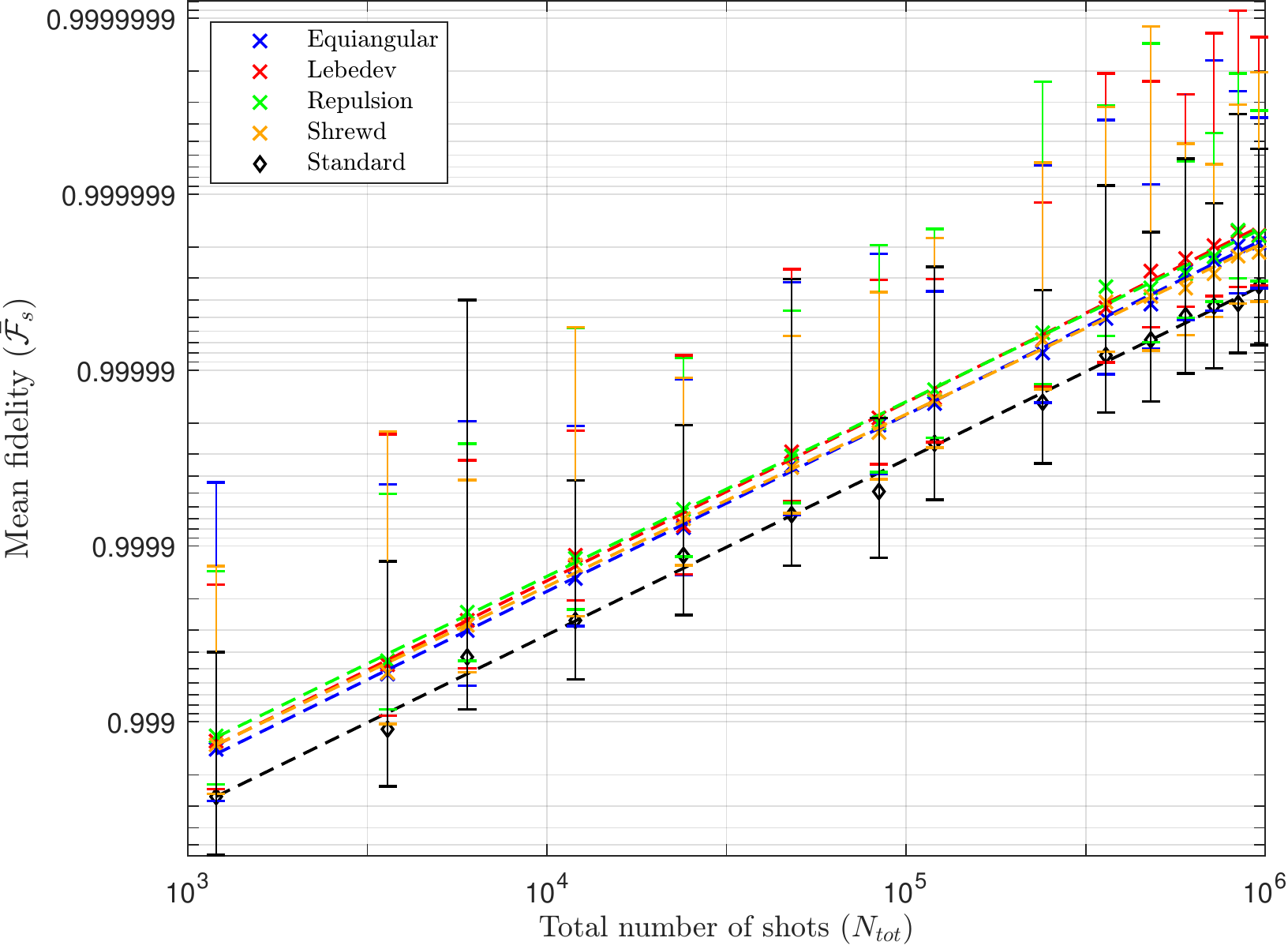}
	\caption{Plot of the mean fidelity ($\bar{\mathcal{F}_s}$) as a function of the total number of shots for different sampling techniques and for the standard state tomography method applied to the quantum state $|\psi\rangle = \frac{1}{\sqrt{2}}(|0\rangle+|1\rangle)$. The mean fidelity is calculated by repeating the simulation 100 times for each data point. In the simulation, only the noise due to a limited number of shots is considered. The standard deviations are shown by the vertical bars for each point.}
	\label{fig:ShotSim_wSD_x}
\end{figure}

Here, we present the plot of the mean fidelities $\bar{\mathcal{F}_s}$ for four different sampling schemes and the standard state tomography method as a function of the total number of shots $N_{tot}$. Fig.~\ref{fig:ShotSim_wSD} shows the same data as in Fig.~\ref{fig:ShotSim} and in addition provides the standard deviations. Fig.~\ref{fig:ShotSim_wSD_x} shows the same numerical study for an equal superposition state $|\psi\rangle = \frac{1}{\sqrt{2}}(|0\rangle+|1\rangle)$. 

\section{Additional figures for visualizing rotation errors}
\label{Sec.:Fig_vis_errors}
\begin{figure}
	\centering
	\includegraphics[scale=1]{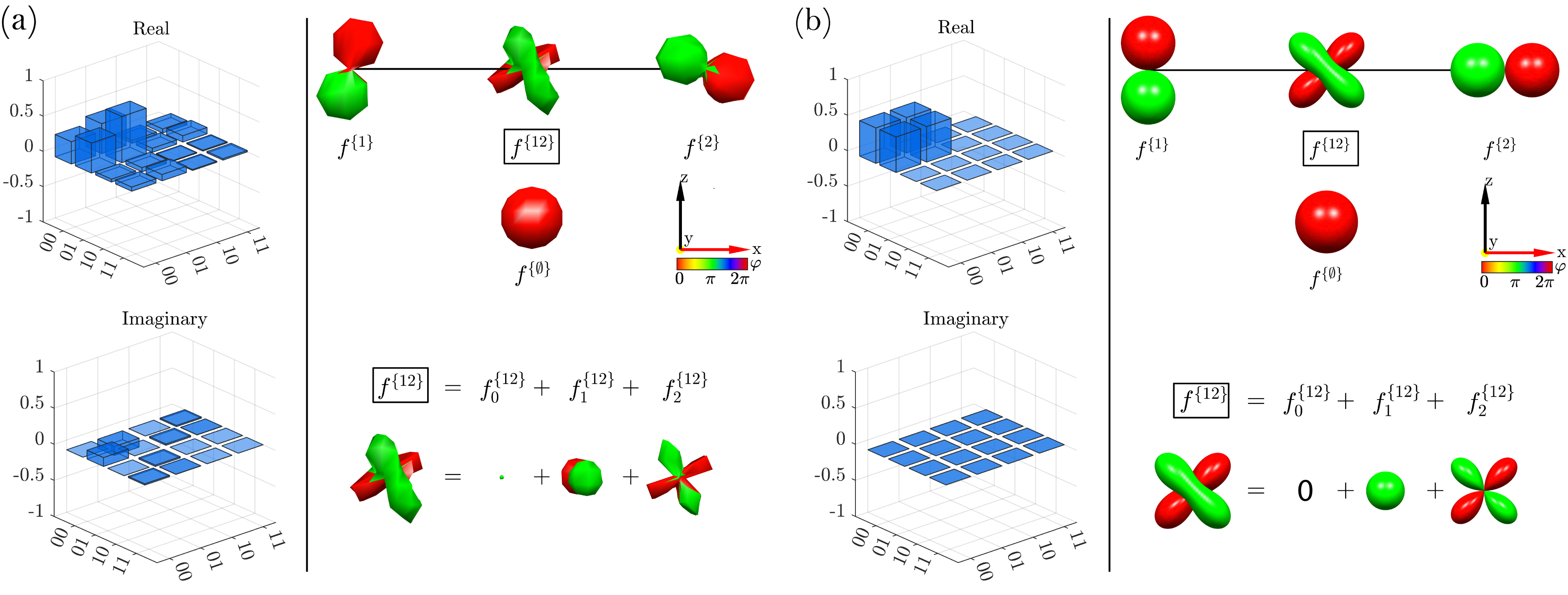}
	\caption{Standard skyscraper and DROPS plot for the state $|\psi\rangle = \frac{1}{\sqrt{2}}(|00\rangle+|01\rangle)$. (a) Experimental tomographed droplets with additional rotations of $\mathrm{U}_3 (\pi/12,0,0)$ on qubit $q_1$, and $\mathrm{U}_3 (\pi/9,\pi/12,0)$ on qubit $q_2$, and (b) simualted droplet plots with no rotation error.  The Figure also shows the respective bilinear droplet function $f^{(12)}$ (box) decomposed into its multipole contribution $f_{j}^{(12)}$ with $j\in\{0,1,2\}$.}
	\label{fig:Error_and_ideal}
\end{figure}

\begin{figure}
	\centering
	\includegraphics[scale=1]{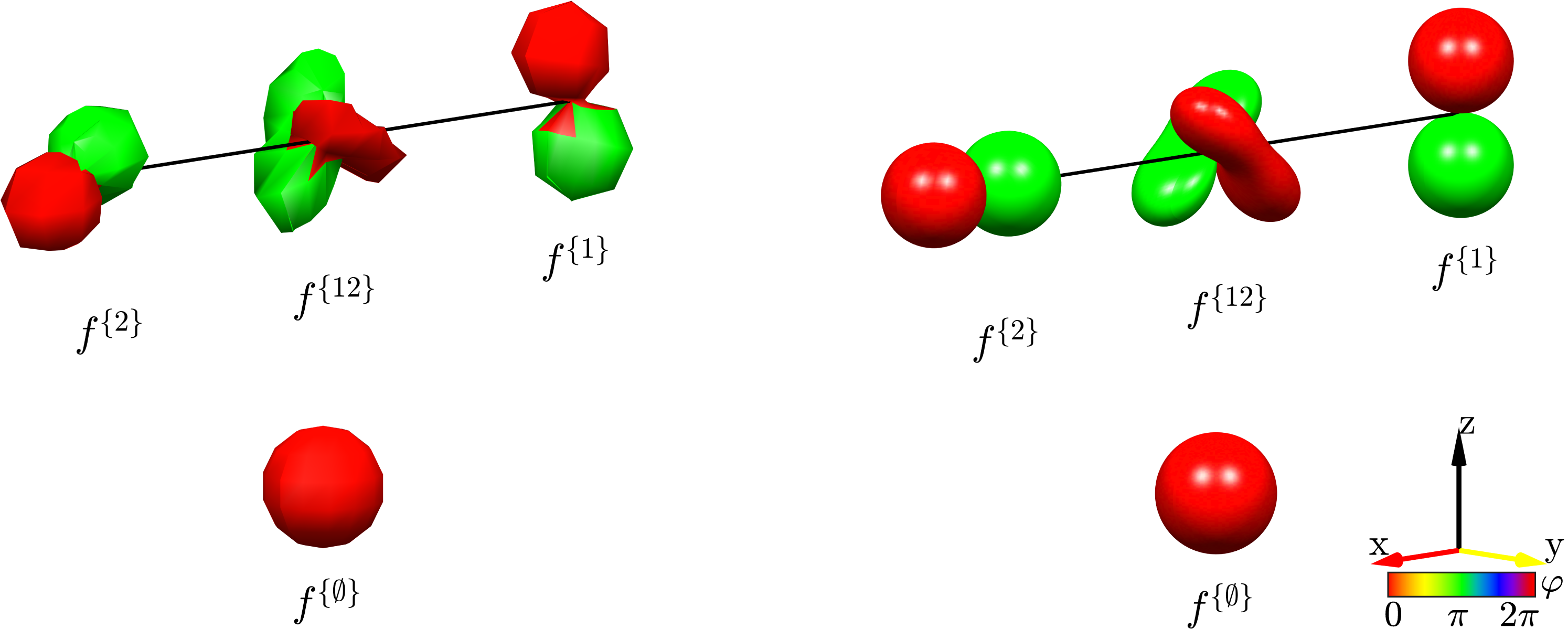}
	\caption{Different perspective of the DROPS visualization of the density matrix corresponding to the state $|\psi\rangle = \frac{1}{\sqrt{2}}(|00\rangle+|01\rangle)$: experimental tomographed droplets with rotation errors (left) and simulated droplets with no errors (right).}
	\label{fig:DiffView}
\end{figure}

For visualizing the result of rotation errors using the DROPS representation, we consider an example in Fig.~\ref{fig:Error} where for conciseness only the skyscraper visualization of the real part of the density matrix is shown. For completeness, in Fig.~\ref{fig:Error_and_ideal} we also show the imaginary part of the skyscraper plot along with decomposed bilinear droplets for the state with and without rotation errors. Fig.~\ref{fig:DiffView} shows the droplet plots from a different perspective to emphasize the misalignment errors. 
\end{document}